\documentclass[journal,twoside]{IEEEtran}
\usepackage{amsfonts}
\usepackage{graphicx}
\usepackage{amsmath}
\usepackage{cite}
\usepackage{url}
\usepackage{array}
\usepackage{color}
\usepackage{amssymb}
\usepackage{epstopdf}
\usepackage{algorithm}
\newcommand{\colvec}[2][.8]{%
  \scalebox{#1}{%
    \renewcommand{\arraystretch}{.8}%
    $\begin{bmatrix}#2\end{bmatrix}$%
  }
}

\DeclareMathOperator{\Tr}{tr}
\newtheorem{theorem}{Theorem}

\newtheorem{lemma}{Lemma}
\newtheorem{remark}{Remark}
\ifCLASSOPTIONcompsoc
\usepackage[caption=false,font=normalsize,labelfont=sf,textfont=sf]{subfig}
\else
\usepackage[caption=false,font=footnotesize]{subfig}
\fi

\begin{document}
\title{Downlink Spectral Efficiency of Cell-Free Massive MIMO Systems with Multi-antenna Users}
\author{Trang C. Mai,~\IEEEmembership{Associate~Member,~IEEE,} Hien Quoc Ngo,~\IEEEmembership{ Member,~IEEE,} and Trung~Q.~Duong,~\IEEEmembership{Senior~Member,~IEEE,} 
\thanks{T. C. Mai, H. Q. Ngo, and T. Q. Duong are with the Institute of Electronics, Communications and Information Technology in Queen's University Belfast, Belfast, U.K. (email: \{trang.mai, hien.ngo, trung.q.duong\}@qub.ac.uk)}
\thanks{The work of Trang C. Mai and Hien Quoc Ngo was supported by the UK Research and Innovation Future Leaders Fellowships under Grant MR/S017666/1. The work of T. Q. Duong was supported in part by the U.K. Royal Academy of Engineering Research Fellowship under Grant RF1415$\backslash$14$\backslash$22.}
\thanks{Parts of this work were presented at the 2018 IEEE GlobalSIP Conf. \cite{Mai2018Cell}}
}
\providecommand{\keywords}[1]{\textbf{\textit{Index terms---}} #1}
\maketitle

\begin{abstract}
This paper studies a cell-free massive multiple-input multiple-output (MIMO) system where its access points (APs) and users are equipped with multiple antennas. Two transmission protocols are considered. In the first transmission protocol, there are no downlink pilots, while in the second transmission protocol, downlink pilots are proposed in order to improve the system performance. In both transmission protocols, the users use the minimum mean-squared error-based successive interference cancellation (MMSE-SIC) scheme to detect the desired signals. For the analysis, we first derive a general spectral efficiency formula with arbitrary side information at the users. Then  analytical expressions for the spectral efficiency of different transmission protocols are derived. To improve the spectral efficiency (SE) of the system, max-min fairness power control (PC) is applied for the first protocol by using the closed-form expression of its SE. Due to the computation complexity of deriving the closed-form performance expression of SE for the second protocol, we apply the optimal power coefficients of the first protocol to the second protocol. Numerical results show that two protocols combining with multi-antenna users are prerequisites to achieve the sub-optimal SE regardless of the number of user in the system.
\end{abstract}

\section{Introduction}
Cellular massive multiple-input multiple-output (MIMO) is currently considered as a key wireless access technology for 5G because it can provide high spectral efficiency (SE) and high energy efficiency (EE) with simple signal processing \cite{ngo2013energy,bjornson2017massive}. In cellular massive MIMO, the BS with massive antenna arrays simultaneously serves all users in its cell on the same time-frequency resource \cite{marzetta2010noncooperative,larsson2013massive,bjornson2015massive,marzetta2016fundamentals}.  

Since cellular massive MIMO is based on cellular topology, its inherent limitation is inter-cell interference. To overcome this limitation, cell-free massive MIMO is introduced \cite{ngo2017cell}. Cell-free massive MIMO can be considered as a useful and scalable version of network MIMO \cite{Shamai2001Enhancing, Venkatesan2007Network} (much in the same way as cellular Massive MIMO
is scalable version of multi-user MIMO). 
In cell-free massive MIMO, a large number of access points (APs), which are geographically distributed over a large area, coherently serve all users on same time-frequency resource \cite{ngo2017cell,interdonato2018ubiquitous}. Cell-free massive MIMO can reap all benefits of massive MIMO (favorable propagation, and channel hardening when using multiple antennas at APs \cite{chen2018Chann}) and network MIMO (increased macro-diversity gain), and hence, it can offer  very high SE, EE, and coverage probability. These benefits can be achieved with simple signal processing and local channel acquisition at each AP.  With the cell-free topology, the excessive handover issue in small-cell systems can be resolved.  Moreover,  poor cell-edge performance, which is typical in small-cell networks, can be resolved in cell-free massive MIMO network by geographically distributing the APs \cite{interdonato2018ubiquitous}. Compared with small cells, cell-free massive MIMO can provide up to a ten-fold improvement in 95\%-likely SE \cite{ngo2017cell,nayebi2017precoding}.
Thus, cell-free massive MIMO has attracted a lot of research interest recently \cite{Ngo2017totalenergy,zhang2017spectral,basharcell,Mai2018Pilot,Mai2019Uplink}. 


Most of previous works exploit the performance of cell-free massive MIMO with single-antenna users. However, in practice, many user's devices of moderate physical size (e.g. laptops, tablets, and smart vehicles) can be equipped with several antennas to increase the multiplexing gain, and to improve system reliability due to the diversity gain. Thus, it is important to evaluate the performance of cell-free massive MIMO with multiple antennas at the users. Moreover, the effect of equipping multiple antennas at the users needs to be well understood to design the systems.  Downlink channel estimation has already investigated in \cite{Ngo2013MU-MIMO}. But \cite{Ngo2013MU-MIMO} considered collocated massive MIMO systems with single-antenna users, orthogonal pilot sequences, and no power control. In  \cite{Interdonato2019disguise,Interdonato2016How}, the authors studied downlink channel estimation of cell-free massive MIMO. But in \cite{Interdonato2019disguise,Interdonato2016How}, each AP has only one antenna, and matched filtering detection is used.

Inspired by the above discussion, in this paper, we analyze the performance of cell-free massive MIMO systems with multiple antennas at both APs and users. In cell-free massive MIMO, each users can be close to several APs, and thus Rician channel model is more reasonable in many scenarios \cite{ozdogan2019performance}. However, in rich scattering environments, the Rayleigh fading model is still reasonable \cite{ngo2017cell,nayebi2017precoding,chen2018Chann,Ngo2017totalenergy,zhang2017spectral,basharcell,Mai2018Pilot,Mai2019Uplink}. In addition, Rayleigh fading model is analytically tractable which helps us to obtain initial and important insights. Therefore, in this paper, we consider cell-free massive MIMO systems using independent, identically distributed (i.i.d.) Rayleigh fading channels. The closed-form expression of downlink SE is derived with taking account of non-orthogonal pilot sequences. As the space between adjacent antennas at the same user is very small, it may causes huge interference to each others. Therefore, in this paper, orthogonal pilot sequences are assigned for antennas at the same user, and those pilot sequences can be reused at antennas of other users. Moreover, the effects of the number of antennas at APs and users on the SE are analyzed through the use of max-min fairness power control. We evaluate the system performance of two protocols for the downlink data transmission of cell-free massive MIMO. In the first protocol, the system operates with two phases: uplink channel estimation and downlink data transmission. As a result, only statistical channel state information (CSI) is available at the users. Note that the statistical CSI depends on large-scale fading which changes very slowly, and it may stay constant for a duration of some 40 small-scale fading coherence intervals \cite{ngo2017cell}. Whereas, in the second protocol, as the level of channel hardening in cell-free massive MIMO is lesser than the one in collocated massive MIMO \cite{chen2018Chann}, we use the downlink channel estimation to improve the system performance. Therefore, in the second protocol, the system operates with three phases: uplink channel estimation, downlink channel estimation and downlink data transmission. As a result, estimated CSI is available at the users. To improve the system performance, both protocols use minimum mean-squared error-based successive interference cancellation (MMSE-SIC) detectors at the users. The computational complexity of MMSE-SIC detectors relates to the inverse operations of the $N\times N$ effective channel gain matrix. Since $N$ is small (several antennas per user), this complexity is low. The second protocol is a generalization of that in previous work on cell-free massive MIMO \cite{Interdonato2016How}, where we consider multi-antennas at users. In both protocols, we also compare system performances using MMSE detectors with the ones using MMSE-SIC detectors. In this paper, to reduce the fronthaul and backhaul requirements, the conjugate beamforming technique is used for both the protocols since it can be implemented in a distributed manner \cite{ngo2017cell}. Other linear processing techniques such as MMSE and zero-forcing (ZF) are better than the conjugate beamforming technique  in terms of the system performance \cite{bjornson2019making}. However, MMSE and ZF need huge fronthaul and backhaul requirements, as we need to send the channel state information to the CPUs, and signal processing is mainly done at the CPUs \cite{bjornson2019making, buzzi2017cell,liu2019spectral}. So it is hard to implement MMSE or ZF in large cell-free massive MIMO networks. Recently, \cite{interdonato2019local} proposed a method which can implement ZF in a distributed manner. But this scheme requires a very large number of antennas at the access points. Thus, cell-free massive MIMO with conjugate beamforming techniques has still received a lot of research attention recently \cite{van2020joint,papazafeiropoulos2020performance,nikbakht2020uplink}.
The main contributions of this paper are as follows.
\begin{itemize}
\item The details of two transmission protocols with and without downlink pilots are presented and analyzed. The channel estimation with non-orthogonal pilot sequences and MMSE-SIC detectors are taken into account.
\item We derive a general formula for the SE with arbitrary side information at the users. Based on this result, analytical expressions for the SE of different transmission protocol are derived. 
\item Max-min fairness power control (PC) is applied for the first protocol to improve the SE of the system. For the second protocol, due to the high complexity associated with computation
for the closed-form of SE, we apply the power control coefficients of the first protocol to the second protocol. Numerical results show that, with those power control coefficients, the SE improves significantly.
\item We investigate the effects of number antennas at both the APs and users.
\item We propose the framework for achieving the sub-optimal system performance regardless of the number of users.
\end{itemize}

The rest of this paper is organized as follows. Section \ref{Sysmol} defines the system model for the downlink cell-free massive MIMO for both
data transmission and channel estimation. Next, Section \ref{SpecEff} derives the achievable downlink SE of cell-free massive MIMO. Then, Section \ref{max-min-pc} derives max-min fairness power control for SE. Section \ref{Numsec}
evaluates the system performance by using numerical results. Finally, the conclusion is drawn in Section \ref{Concl}.

\textit{Notation}: The superscripts $()^\ast$, $()^T$, and $()^H$ stand for the conjugate, transpose, and conjugate-transpose, respectively. The Euclidean norm, the expectation operators, and the determinant of matrix are denoted by $\|\cdot \|$, $\mathbb {E}\left \{{\cdot }\right \}$, and $|.|$, respectively. $\text{var}(.)$ denotes variance. In addition, ${z} \sim \mathcal {CN}\left ({{0},{\sigma ^{2}}}\right )$ denotes a circularly symmetric complex Gaussian random variable (RV) $z$ with zero mean and variance $\sigma ^{2}$, and $\mathbb {C}^{L \times N}$ denotes the ${L \times N}$ matrix. Finally, ${z} \sim \mathcal {N}(0,\sigma ^{2})$ denotes a real-valued Gaussian RV.
\section{System Model}\label{Sysmol}
We consider a cell-free massive MIMO system operating in time division duplex (TDD) mode with $M$ APs and $K$ users randomly located within a large area. Each AP has $L$ antennas, whereas each user has $N$ antennas. 
Let $\textbf{G}_{mk} \in \mathbb {C}^{L \times N}$ be the channel response matrix between the $k$-th user and the $m$-th AP. Then, 
\begin{equation}
\textbf{G}_{mk} = \beta^{1/2}_{mk}\textbf{H}_{mk},
\end{equation}
where $\beta_{mk}$ is large-scale fading between the $k$-th user and the $m$-th AP, and $\textbf{H}_{mk}$ is the $L\times N$ small-scale fading matrix whose elements are assumed to be independent and identically distributed (i.i.d.) $\mathcal {CN}\left ({{0},{1}}\right )$ RVs. In this work, we focus on the downlink transmission, and hence, the uplink data transmission is neglected. Specifically, we consider two transmission protocols. The first protocol has two phases: the uplink channel estimation and the downlink data transmission. The second protocol has three phases: the uplink channel estimation, the downlink channel estimation and the downlink data transmission. System model of the two protocols will be presented in detail in the rest of this section. 

\subsection{Transmission Protocol 1 - No Downlink Pilots}
This transmission protocol is commonly used in previous studies of cell-free massive MIMO systems. Each user relies on the channel hardening property of massive MIMO technology to detect the desired signals. So there is no downlink channel estimation phase \cite{ngo2017cell}.
\subsubsection{Uplink Channel Estimation}\label{UCE_P1}
In this phase, all users will send pilot signals to the APs. Then each AP will estimate its channels to all users using the received pilot signals. Let $\tau_{\text{u}}$ be the length of the uplink training duration per coherence interval, and $\pmb{\Phi}_{\text{u},k} \in \mathbb {C}^{\tau_{\text{u}} \times N}$, where its $n$-th column satisfies  $\|\pmb{\phi}_{\text{u},k,n} \| = 1$, $\forall n \in N$, be a pilot matrix of the $k$-th user.
Then, the received signal at the $m$-th AP is
\begin{equation} 
\textbf{Y}_{\text{u}, m}=\sum_{k=1}^{K}\sqrt{\tau_{\text{u}}\rho_{\text{u}}}\textbf{G}_{mk}\pmb{\Phi}_{\text{u},k}^H+\textbf{W}_{\text{u},m}, 
\end{equation}
where $\rho_{\text{u}}$ is the normalized signal-to-noise ratio (SNR) of each uplink pilot symbol and $\textbf{W}_{\text{u},m}$ is the $L\times\tau_{\text{u}}$ matrix of additive noise at the $m$-th AP. We assume that the elements of $\textbf{W}_{\text{u}, m}$ are i.i.d. $\mathcal {CN}\left ({{0},{1}}\right )$ RVs. A projection of $\textbf{Y}_{\text{u}, m}$ onto $\pmb{\Phi}_{\text{u}, k}$ is
\begin{align} 
\textbf{Y}_{\text{u},mk} &=\textbf{Y}_{\text{u},m}\pmb{\Phi}_{\text{u},k}\notag \\ 
 &= \sum_{i=1}^{K}\sqrt{\tau_{\text{u}}\rho_{\text{u}}}\textbf{G}_{mi}\pmb{\Phi}_{\text{u},ik} + \textbf{W}_{\text{u},mk},
\end{align}
where $\pmb{\Phi}_{\text{u},ik} \triangleq \pmb{\Phi}_{\text{u},i}^H\pmb{\Phi}_{\text{u},k}$, and $\textbf{W}_{\text{u},mk} \triangleq \textbf{W}_{\text{u},m}\pmb{\Phi}_{\text{u},k}$.
By stacking all columns of $\textbf{Y}_{\text{u},mk}$ on top of each other, we have
\begin{align} 
\text{vec}(\textbf{Y}_{\text{u},mk}) &= \sqrt{\tau_{\text{u}}\rho_{\text{u}}}\sum_{i=1}^{K}(\pmb{\Phi}_{\text{u},ik}^T \otimes \textbf{I}_L) \text{vec}(\textbf{G}_{mi})+ \text{vec}(\textbf{W}_{\text{u},mk})\notag\\
&=\sqrt{\tau_{\text{u}}\rho_{\text{u}}}\sum_{i=1}^{K}\tilde{\pmb{\Phi}}_{ik} \text{vec}(\textbf{G}_{mi})+ \text{vec}(\textbf{W}_{\text{u},mk}),
\end{align}
where vec(.) is the vectorization operation, and $\tilde{\pmb{\Phi}}_{\text{u},ik} \triangleq \pmb{\Phi}_{\text{u},ik}^T \otimes \textbf{I}_L$. Since the distance between adjacent antennas at the same user is very small, non-orthogonal pilots may cause huge interference to each others. To mitigate interference between antennas of the same user, we assume that orthogonal pilot sequences are assigned for antennas of each user, but these pilot sequences can be
reused in other users. Then, MMSE estimation of $\text{vec}(\textbf{G}_{mk})$ given $\text{vec}(\textbf{Y}_{\text{u},mk})$ is expressed by \cite{kay1993fundamentals}
\begin{align}\label{vecchann} 
\text{vec}(\hat{\textbf{G}}_{mk})&= \sqrt{\tau_{\text{u}}\rho_{\text{u}}}\beta_{mk}\left(\tau_{\text{u}}\rho_{\text{u}}\sum_{i=1}^{K}\tilde{\pmb{\Phi}}_{\text{u},ik}\beta_{mi}\tilde{\pmb{\Phi}}^H_{\text{u},ik}+ \textbf{I}_{LN}  \right)^{-1}\notag\\
&\quad\times\text{vec}(\textbf{Y}_{\text{u},mk}).
\end{align}
\begin{lemma}\label{lemma1} 
The estimate of the channel matrix $\textbf{G}_{mk}$ is 
\begin{align}\label{estchan1}
\hat{\textbf{G}}_{mk} = \textbf{Y}_{\text{u},mk} \textbf{A}_{mk}, 
\end{align}
where 
\begin{equation} 
\textbf{A}_{mk} \triangleq \sqrt{\tau_{\text{u}}\rho_{\text{u}}}\beta_{mk}\left(\tau_{\text{u}}\rho_{\text{u}}\sum_{i=1}^{K}\beta_{mi}\pmb{\Phi}^H_{\text{u},ik}\pmb{\Phi}_{\text{u},ik}+ \textbf{I}_N\right)^{-1}.
\end{equation}
\end{lemma}
\begin{IEEEproof}
	See Appendix \ref{AppendixA}.
\end{IEEEproof}

\subsubsection{Downlink data transmission}\label{DDT_P1}
In this phase, each AP uses the channel estimates in the uplink channel estimation phase together with the conventional conjugate beamforming technique to precode the desired symbols \cite{ngo2017cell}. Then the precoded signal will be sent to all users. The $L\times 1$ transmitted  signal from the $m$-th AP is 
\begin{equation} 
\textbf{x}_m= \sqrt{\rho}\sum_{k=1}^{K}\eta^{1/2}_{mk}\hat{\textbf{G}}_{mk}\textbf{q}_{k} ,
\end{equation}
where $\textbf{q}_{k}$, with $ \mathbb {E} \{\textbf{q}_{k}\textbf{q}^{H}_{k} \} = \textbf{I}_{N}$, is the vector of symbols intended for the $k$-th user, $\rho$ is the normalized transmit SNR constraint at the $m$-th AP, and $\eta_{mk}$ is power control coefficient corresponding to the $k$-th user.
The power control coefficients $\eta_{mk}$ are chosen to satisfy the power constraint at each AP:  $\text{E}\{\Arrowvert \textbf{x}_m \Arrowvert^2\}\le \rho$ which is equivalent to 
\begin{equation}\label{powcontraint} 
\sqrt{\tau_{\text{u}}\rho_{\text{u}}}\sum_{k=1}^{K}\eta_{mk}\beta_{mk}\Tr \left(\textbf{A}_{mk}\right) \le \frac{1}{L}.
\end{equation}
The received signal at the $k$-th user is
\begin{align}\label{data_recei_eq} 
\textbf{r}_k &= \sum_{m=1}^{M}\textbf{G}^H_{mk}\textbf{x}_m + \textbf{n}_k\notag\\
&= \sqrt{\rho}\sum_{k'=1}^{K}\textbf{D}_{kk'}\textbf{q}_{k'} + \textbf{n}_k, 
\end{align}
where $\textbf{D}_{kk'} \triangleq \sum_{m=1}^{M}\eta_{mk'}^{1/2}\textbf{G}^H_{mk}\hat{\textbf{G}}_{mk'}$ denotes the effective downlink channel for the $k'$-th user and $\textbf{n}_k$ is the noise vector. The elements of $\textbf{n}_k$ are assumed to be i.i.d. $\mathcal {CN}\left ({{0},{1}}\right )$.

\subsection{Transmission Protocol 2 - With Downlink Pilots}
Different from cellular massive MIMO, cell-free massive MIMO offers lesser channel hardening. Therefore it is good to estimate channel at the users via the downlink pilots. Here for the analysis simplicity, we assume that orthogonal pilot sequences are used for the downlink channel estimation phase.
\subsubsection{Uplink Channel Estimation}
This phase is the same as that of Transmission Protocol 1. See Section~\ref{UCE_P1}.

\subsubsection{Downlink Channel Estimation}
From the received signal (\ref{data_recei_eq}), to detect the desired signal, each user does not need to estimate all channel matrices $\textbf{G}_{mk}$. Instead, it needs to estimate only the effective channel gain matrices $\textbf{D}_{kk'}$ which have much lower dimension. To do this, the pilot sequences will be precoded before being sent to all users \cite{Interdonato2016How}.
Let $\tau_{\text{d}}$ be the length of the downlink training duration per coherence interval, and $\pmb{\Phi}_{\text{d},k} \in \mathbb {C}^{\tau_{\text{d}} \times N}$ be a pilot matrix for the $k$-th user that satisfies
\begin{equation*}
\pmb{\Phi}^H_{\text{d},k} \pmb{\Phi}_{\text{d},k'} =
\begin{cases}
\textbf{I}_N & \text{if } k = k',\\
0 & \text{if } k \neq k'.
\end{cases}
\end{equation*}
Then, precoded pilot matrix which is transmitted from the $m$-th AP is
\begin{equation} 
\textbf{X}_{\text{d}, m}=\sqrt{\tau_{\text{d}}\rho_{\text{d}}}\sum_{k'=1}^{K}\eta_{mk'}^{1/2}\hat{\textbf{G}}_{mk'}\pmb{\Phi}_{\text{d},k'}^H,
\end{equation}
where $\rho_{\text{d}}$ is the normalized signal-to-noise ratio (SNR) of the $m$-th AP. Then, the received pilot signal at the $k$-th user is 
\begin{align} 
\textbf{Y}_{\text{d},k} &=\sqrt{\tau_{\text{d}}\rho_{\text{d}}}\sum_{m=1}^{M}\sum_{k'=1}^{K}\eta_{mk'}^{1/2}\textbf{G}^H_{mk}\hat{\textbf{G}}_{mk'}\pmb{\Phi}_{\text{d},k'}^H+\textbf{W}_{\text{d},k},\notag \\ 
&=\sqrt{\tau_{\text{d}}\rho_{\text{d}}}\sum_{k'=1}^{K}\textbf{D}_{kk'}\pmb{\Phi}_{\text{d},k'}^H + \textbf{W}_{\text{d},k}.
\end{align}
\begin{lemma}\label{lemma2} 
MMSE estimation of $\textbf{D}_{kk'}$, $\forall k, k' = 1, 2,\hdots, K$, given $\textbf{Y}_{\text{dp},k}$ is $\hat{\textbf{D}}_{kk'}$, whose $ij$-th element is
\begin{align}\label{eq_d_kkface}
\hat{d}_{kk',ij}=\begin{cases}
\frac{\sqrt{\tau_{\text{d}}\rho_{\text{d}}}\left(\xi_{kk,i} + \kappa_{kk,i}^2 \right) y_{\text{d},k,ij}  + \kappa_{kk,i}}{\tau_{\text{d}}\rho_{\text{d}}\left(\xi_{kk,i} + \kappa_{kk,i}^2 \right) + 1}  & \text{if } k = k', i = j, \\
\frac{\sqrt{\tau_{\text{d}}\rho_{\text{d}}}\xi_{kk',j}y_{\text{d},k,ij} }{\tau_{\text{d}}\rho_{\text{d}}\xi_{kk',j} + 1} & \text{otherwise},
\end{cases}
\end{align}
where $y_{\text{d},k,ij}$ is $ij$-element of the matrix $\textbf{Y}_{\text{dp},k}$, $\xi_{kk,i} = L\sum^M_{m=1}\eta_{mk}\beta_{mk}\gamma_{mk,i}$, $\gamma_{mk,i} = \bigg[ \tau_{\text{u}}\rho_{\text{u}}\beta^2_{mk} \left(\tau_{\text{u}}\rho_{\text{u}}\sum_{i=1}^{K}\tilde{\pmb{\Phi}}_{\text{u},ik}\beta_{mi}\tilde{\pmb{\Phi}}^H_{\text{u},ik}+ \textbf{I}_{LN}  \right)^{-1} \bigg ]_{(i-1)L + l}$, $\kappa_{kk,i} = L\sum_{m=1}^{M}\eta^{1/2}_{mk}\gamma_{mk,i} $, and $\xi_{kk',i} = L\sum^M_{m=1}\eta_{mk'}\beta_{mk}\gamma_{mk',i}$.
\end{lemma}
\begin{IEEEproof}
	See Appendix \ref{AppendixB}.
\end{IEEEproof}
Note that $\textbf{D}_{kk'}$ is an  $N \times N$ matrix. Since $N$ is small, the corresponding complexity of the MMSE estimation is low. 
\subsubsection{Downlink data transmission}
The downlink transmission of this protocol is the same as that of Protocol 1, see Section \ref{DDT_P1}. But in this protocol, since each user estimates the effective channel gain matrices, it will use this information to detect the desired symbols.
\section{Spectral Efficiency}\label{SpecEff}
In this section, we derive analytical expressions for the SEs of transmission protocols 1 and 2 assuming that each user uses the MMSE-SIC scheme to detect the desired symbols. The SE under the assumption that the users have perfect CSI is derived as a benchmark. In addition, for the comparison,  we derive the SE expression for case that the users use the simple linear MMSE detector instead of the MMSE-SIC detector. Note that the computational complexity of MMSE-SIC detectors which relates to the inverse operations of the $N\times N$ effective channel gain matrix, is low. Since different SEs correspond to the different of side information available at the users, we first provide a general SE expression with side information at the users as in the following theorem.
\begin{theorem}\label{theo_SE}
The achievable downlink SE of the $k$-th user with MMSE-SIC detection scheme given the received signal $\textbf{r}_{k}$ in (\ref{data_recei_eq}) and side information $\mathbf{\Theta}_{k}$ (assuming that $\mathbf{\Theta}_{k}$ is independent of $\textbf{q}_{k}$) is given by
\begin{align}\label{R_k1}  
R_k = (1 - \tau_{\text{tot}}/\tau_{\text{c}})\mathbb {E}\left\{ \log_{2} \left| \textbf{I}_{N} + \mathbf{\Upsilon}^a_{kk} \right|\right\},
\end{align}
where $\tau_{\text{tot}}$ is the total training duration per coherence interval $\tau_{\text{c}}$,
$\mathbf{\Upsilon}^a_{kk} \triangleq \rho \mathbb {E}\{\textbf{D}^H_{kk}|\mathbf{\Theta}_{k}\} \left(\mathbf{\Psi}^a_{kk} \right)^{-1} \mathbb {E}\{\textbf{D}_{kk}|\mathbf{\Theta}_{k}\}$,
and $\mathbf{\Psi}^a_{kk} \triangleq \textbf{I}_{N} + \mathbb {E}\{(\rho\sum^{K}_{k'=1}\textbf{D}_{kk'}\textbf{D}^H_{kk'}|\mathbf{\Theta}_{k})\} - \rho \mathbb {E}\{\textbf{D}_{kk}|\mathbf{\Theta}_{k}\} \mathbb {E}\{\textbf{D}^H_{kk}|\mathbf{\Theta}_{k}\}.$

\end{theorem}
\begin{IEEEproof}
	 See Appendix \ref{Appendix_theo_SE}.
\end{IEEEproof}
\subsection{Achievable Downlink SE for Protocol 1}\label{Scheme1}
For this transmission procotol, there are no downlink pilots. Each user uses ony the statistic CSI for signal detection. This means that  $\mathbf{\Theta}_{k} = \bar{\textbf{D}}_{kk} \triangleq \mathbb {E}\{\textbf{D}_{kk}\}$.
With $\mathbf{\Theta}_{k} = \bar{\textbf{D}}_{kk}$, the achievable downlink SE of the $k$-th user in (\ref{R_k1}) becomes
\begin{align}\label{R_k_statistic}  
R^{\text{St-SIC}}_k = (1 - \tau_{\text{u}}/\tau_c) \log_{2} \left| \textbf{I}_{N} + \rho\bar{\textbf{D}}^H_{kk} \left(\mathbf{\Psi}^b_{kk}\right)^{-1} \bar{\textbf{D}}_{kk} \right|,
\end{align}
where 
\begin{align}
\mathbf{\Psi}^b_{kk} = \textbf{I}_{N} + \mathbb {E}\left\{\left(\rho\sum^{K}_{k'=1}\textbf{D}_{kk'}\textbf{D}^H_{kk'}\right)\right\} - \rho\bar{\textbf{D}}_{kk}\bar{\textbf{D}}^H_{kk}. \notag
\end{align}
By deriving all expectations, we obtain the closed-form expression for the achievable SE (\ref{muin1}) as in the following theorem.
\begin{theorem}\label{theo_closedform}
Given statistic CSI, says $\mathbf{\Theta}_{k} = \bar{\textbf{D}}_{kk}$, and using MMSE-SIC detectors, the achievable downlink SE for the $k$-th user can be represented
in closed-form as
\begin{align}\label{muin1} 
R^{\text{St-SIC}}_k = (1 - \tau_{\text{u}}/\tau_c) \log_{2} \left| \textbf{I}_{N} + \rho\bar{\textbf{D}}^H_{kk} \left(\mathbf{\Psi}^b_{kk}\right)^{-1} \bar{\textbf{D}}_{kk} \right|,
\end{align}
where
\begin{align} 
\bar{\textbf{D}}_{kk}=L\sqrt{\tau_{\text{u}}\rho_{\text{u}}}\sum_{m=1}^{M}\eta^{1/2}_{mk}\beta_{mk}\textbf{A}_{mk},
\end{align} 
and
\begin{align}\label{seq} 
&\mathbf{\Psi}^b_{kk}\notag\\
&= L\tau_{\text{u}}\rho_{\text{u}}\rho\sum_{m=1}^{M}\Bigg(\sum_{k'=1}^{K}\eta_{mk'}\beta^2_{mk}\textbf{C}_{mkk'} - L\eta_{mk}\beta^2_{mk}\textbf{A}_{mk}\textbf{A}^H_{mk} \notag\\ 
&\quad + L\sum_{n \ne m}^{M}\sum_{k'\ne k}^{K}\eta^{1/2}_{mk'}\eta^{1/2}_{nk'}\beta_{mk}\beta_{nk} \Phi_{kk'}\textbf{A}_{mk'}\textbf{A}^H_{nk'}\Phi^H_{kk'}\notag\\
&\quad +\sum_{k'=1}^{K}\sum_{i \ne k}^{K}\eta_{mk'}\beta_{mk}\beta_{mi}\Tr(\Phi_{ik'}\textbf{A}_{mk'}\textbf{A}^H_{mk'}\Phi^H_{ik'})\textbf{I}_{N} \Bigg) \notag\\ 
&\quad+ L\rho\sum_{m=1}^{M}\sum_{k'=1}^{K}\beta_{mk}\eta_{mk'}\Tr(\textbf{A}_{mk'}\textbf{A}^H_{mk'})\textbf{I}_{N} + \textbf{I}_{N},
\end{align}
where $\textbf{C}_{mkk'}$ is a $N \times N$ diagonal matrix that $\left[\textbf{C}_{mkk'}\right]_{ii} = \sum_{n = 1}^{N} b_{mkk',nn} + Lb_{mkk',ii}$ with $b_{mkk',ii} \triangleq \left[\textbf{B}_{mkk'}\right]_{ii}= \left[\Phi_{kk'}\textbf{A}_{mk'}\textbf{A}^H_{mk'}\Phi^H_{kk'}\right]_{ii}$.
\end{theorem}
\begin{IEEEproof}
	See Appendix \ref{AppendixE}.
\end{IEEEproof}
\begin{remark}
Note that, a minor correction is updated in (\ref{seq}) when comparing with that in conference version in \cite{Mai2018Cell}.  
\end{remark}
The lower bound of the downlink SE in Theorem~\ref{theo_closedform} can be achieved by using per-user-basis MMSE-SIC detector while treating co-user interference plus noise as uncorrelated Gaussian noise with the assumption that $\textbf{q}_{k} \thicksim \mathcal {CN}\left ({{0},{\textbf{I}_{N}}}\right )$ .
\begin{remark}
In the special case that all APs and users have a single antenna, i.e., $L=N=1$, the spectral efficiency (\ref{muin1}) is identical to the one in \cite{ngo2017cell}.
\end{remark}

\subsection{Achievable Downlink SE for Protocol 2}\label{Scheme2}
For this transmission protocol, each user acquires the estimates of the effective channel gains via the downlink pilots. More precisely, we have $\mathbf{\Theta}_{k} = \{\hat{\textbf{D}}_{ki}\}$, $\forall i \in K$.
Using Theorem 1, we obtain the following achievable SE:
\begin{align}\label{se_sch2}  
R_k = \left(1 - \frac{\tau_{\text{u}} + \tau_{\text{d}}}{\tau_{\text{c}}}\right)\mathbb {E}\left\{ \log_{2} \left| \textbf{I}_{N} + \mathbf{\Upsilon}^c_{kk} \right|\right\},
\end{align}
where 
$\mathbf{\Upsilon}^c_{kk} \triangleq \rho \mathbb {E}\left\{\textbf{D}^H_{kk}|\{\hat{\textbf{D}}_{ki}\}\right\} \left(\mathbf{\Psi}^c_{kk}\right)^{-1}  \mathbb {E}\left\{\textbf{D}_{kk}|\{\hat{\textbf{D}}_{ki}\}\right\}$,
and $\mathbf{\Psi}^c_{kk} \triangleq \textbf{I}_{N} + \mathbb {E}\left\{(\rho\sum^{K}_{k'=1}\textbf{D}_{kk'}\textbf{D}^H_{kk'}|\{\hat{\textbf{D}}_{ki}\})\right\} - \rho \mathbb {E}\left\{\textbf{D}_{kk}|\{\hat{\textbf{D}}_{ki}\}\right\} \mathbb {E}\left\{\textbf{D}^H_{kk}|\{\hat{\textbf{D}}_{ki}\}\right\}.$ Since the elements of $\textbf{D}_{kk'}$ are not Gaussian distributed, the elements of their MMSE estimate $\hat{\textbf{D}}_{kk'}$ and corresponding elements of estimation error $\tilde{\textbf{D}}_{kk'}$ are uncorrelated, but not independent. This makes (\ref{se_sch2}) hard to be computed in closed form. However,  elements of $\textbf{D}_{kk'}$  are very close to Gaussian, especially when $M$ is large. This is shown in the Lemma \ref{lemma3} as follows.
\begin{lemma}\label{lemma3} 
The elements of downlink effective channels $\textbf{D}_{kk'}, (\forall k, k' = 1,2, \ldots, K)$ converge in distribution to Gaussian distribution as follows:
\begin{align} 
&d_{kk,ij}\overset{d}{\rightarrow} \mathcal{CN}(0, \xi_{kk,j}),\text{as}\ M\rightarrow\infty, \text{ and } \forall i \neq j,\notag\\
&d_{kk,ii}\overset{d}{\rightarrow} \mathcal{N}\left(L\sum_{m=1}^{M}\sqrt{\eta_{mk}}\gamma_{mk,i},L\sum_{m=1}^{M}\eta_{mk}\gamma_{mk,i}^{2}\right),\notag\\
&\quad\quad\quad\quad \text{as}\ M\rightarrow\infty,\text{ and}\notag\\
&d_{kk',ij}\overset{d}{\rightarrow} \mathcal{CN}(0, \xi_{kk',i}),\text{as}\ M\rightarrow\infty,\forall k' \ne k,  \text{ and } \forall i, j \in N, \notag
\end{align}
where $\overset{d}{\rightarrow}$ denotes convergence in distribution.
\end{lemma}
\begin{IEEEproof}
	See Appendix \ref{Appendix_proflemma3}. 
\end{IEEEproof}
\begin{figure}[!h]
\centering
\includegraphics[width=0.45\textwidth]{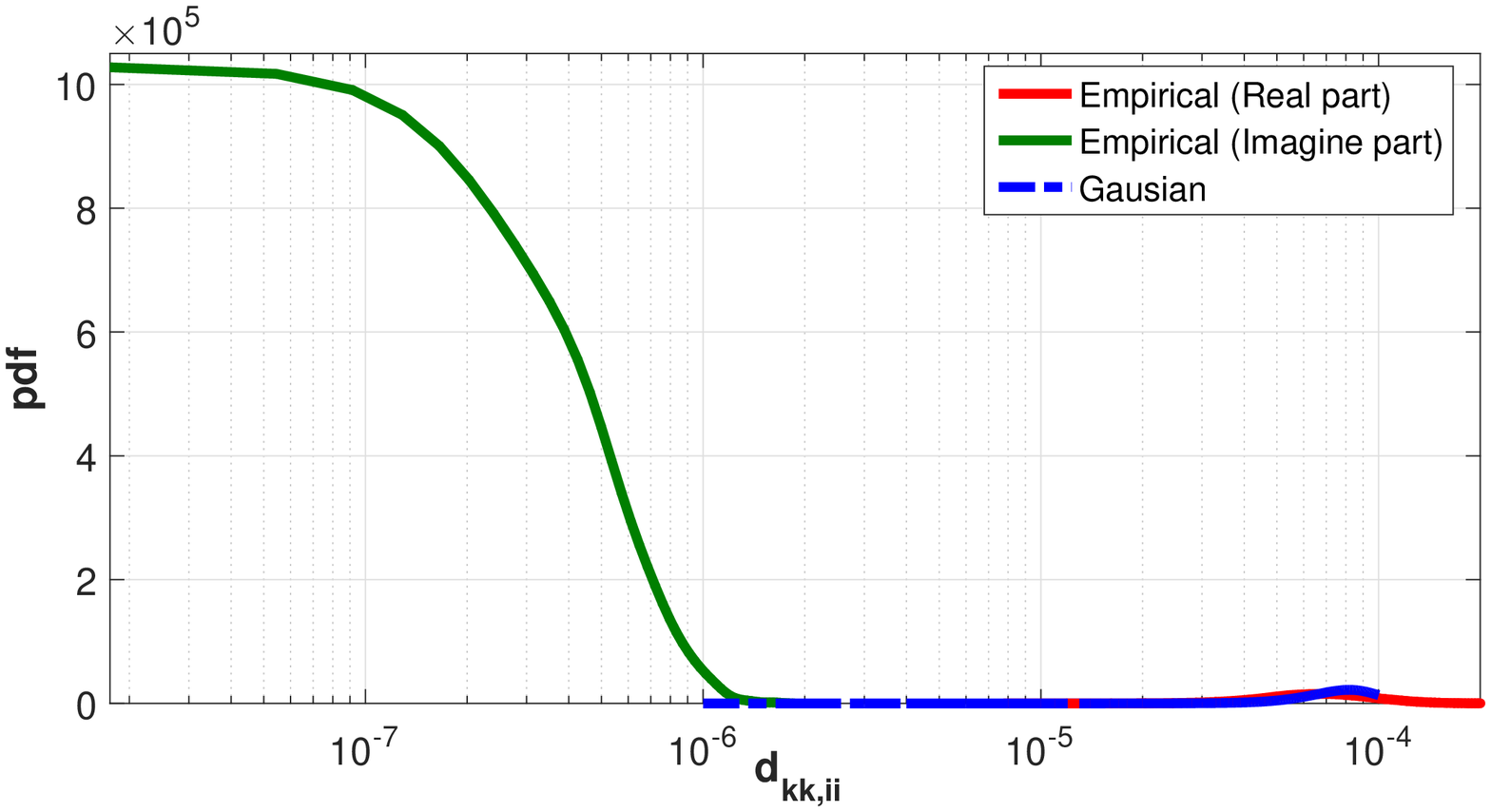}\\
\includegraphics[width=0.23\textwidth]{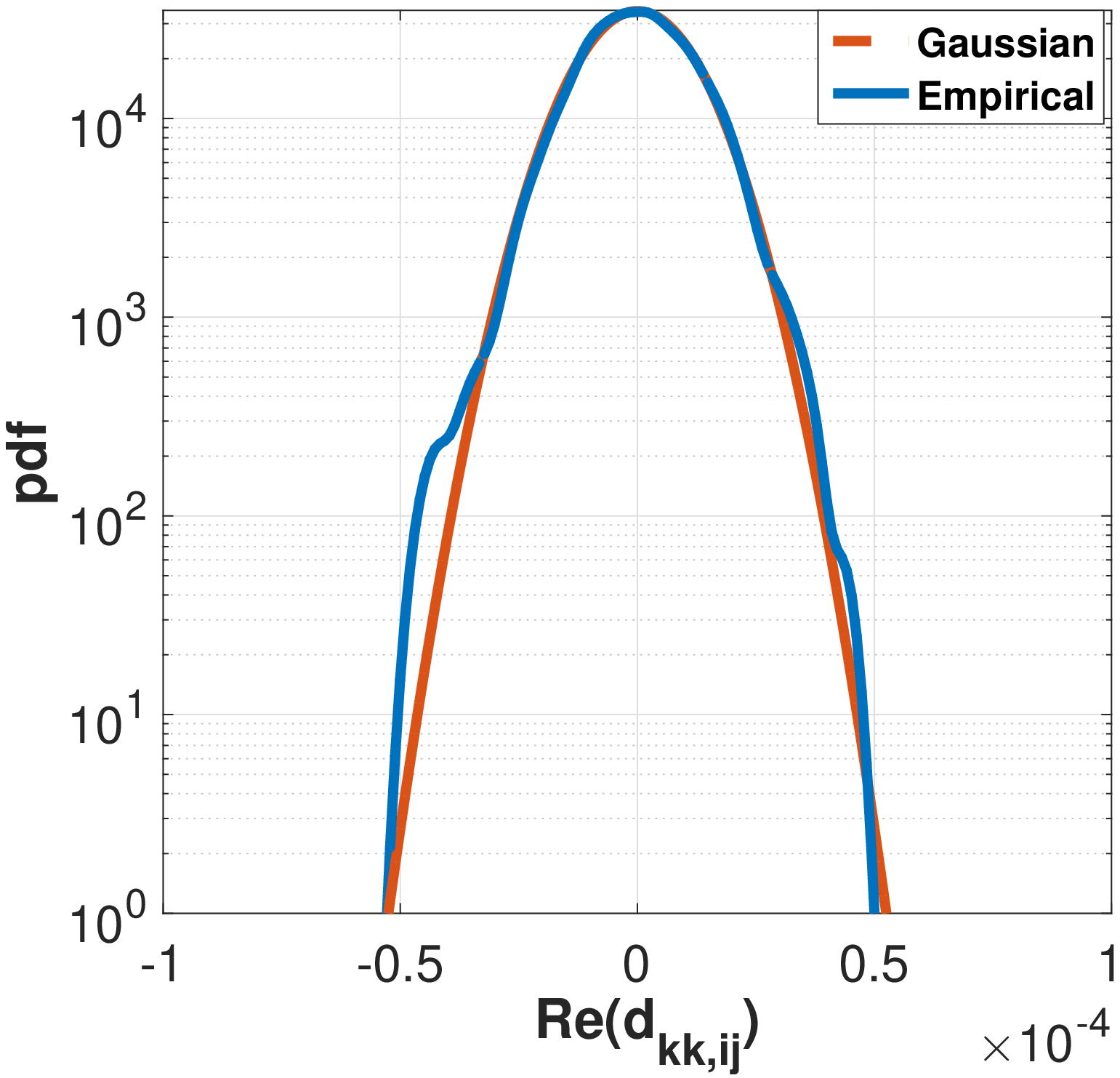}
\includegraphics[width=0.23\textwidth]{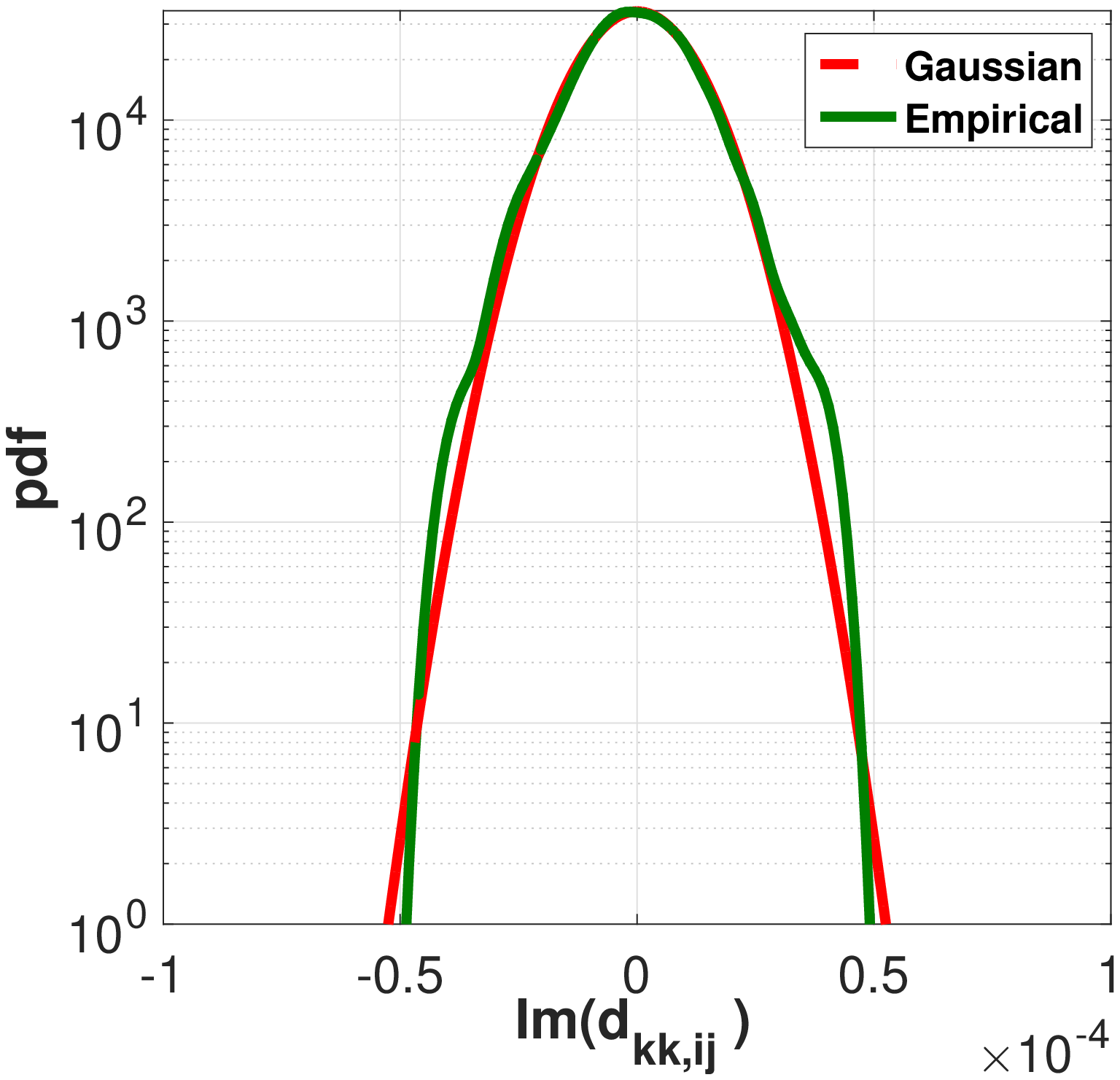}
\caption{True (Empirical) v.s. approximated (Gaussian) pdf of $d_{kk,ij})$, $\forall i,j = 1, 2, \ldots, N$.}
\label{d_kk}
\end{figure}

\begin{figure}[!h]
\centering
\includegraphics[width=0.23\textwidth]{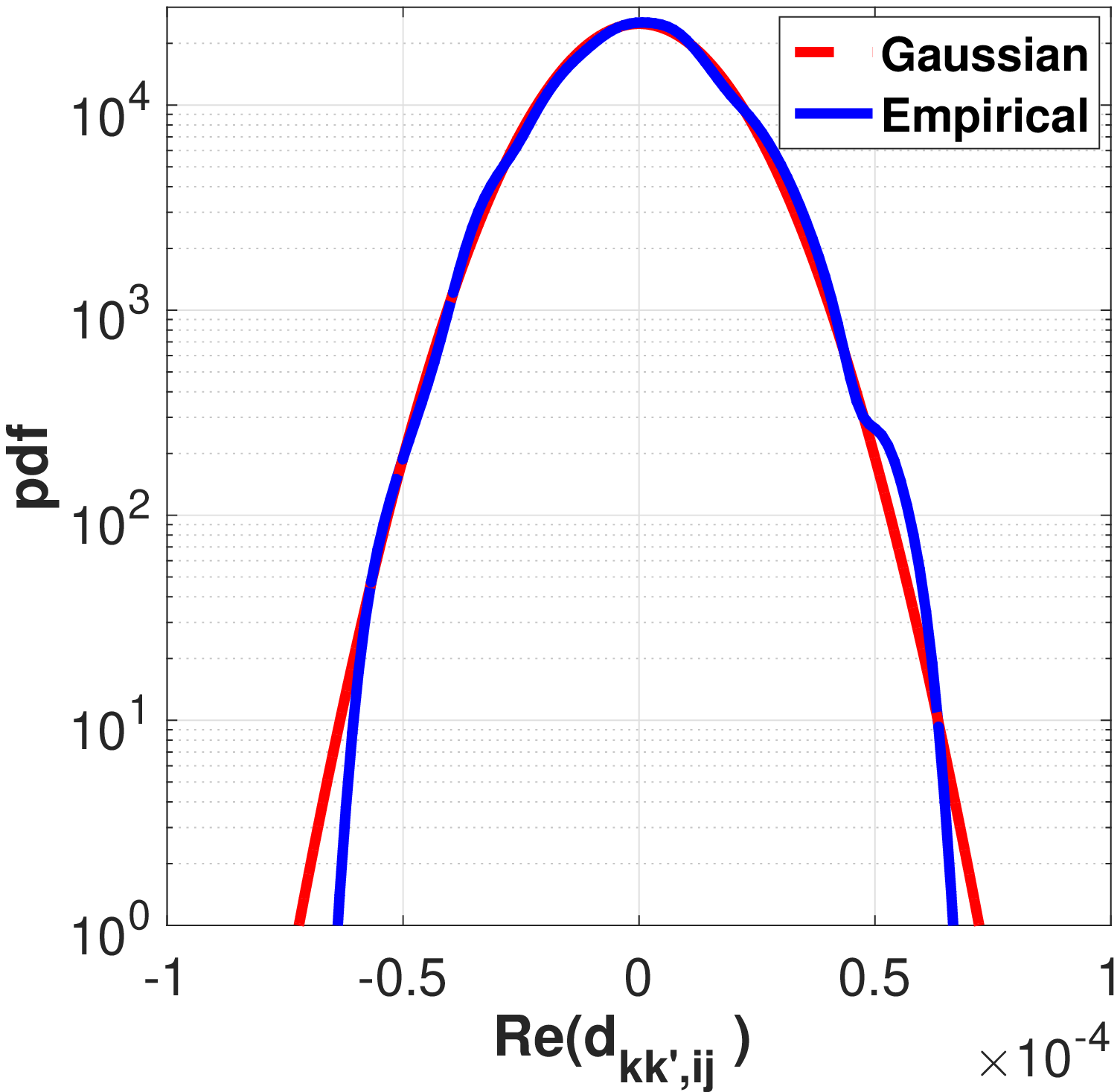}
\includegraphics[width=0.23\textwidth]{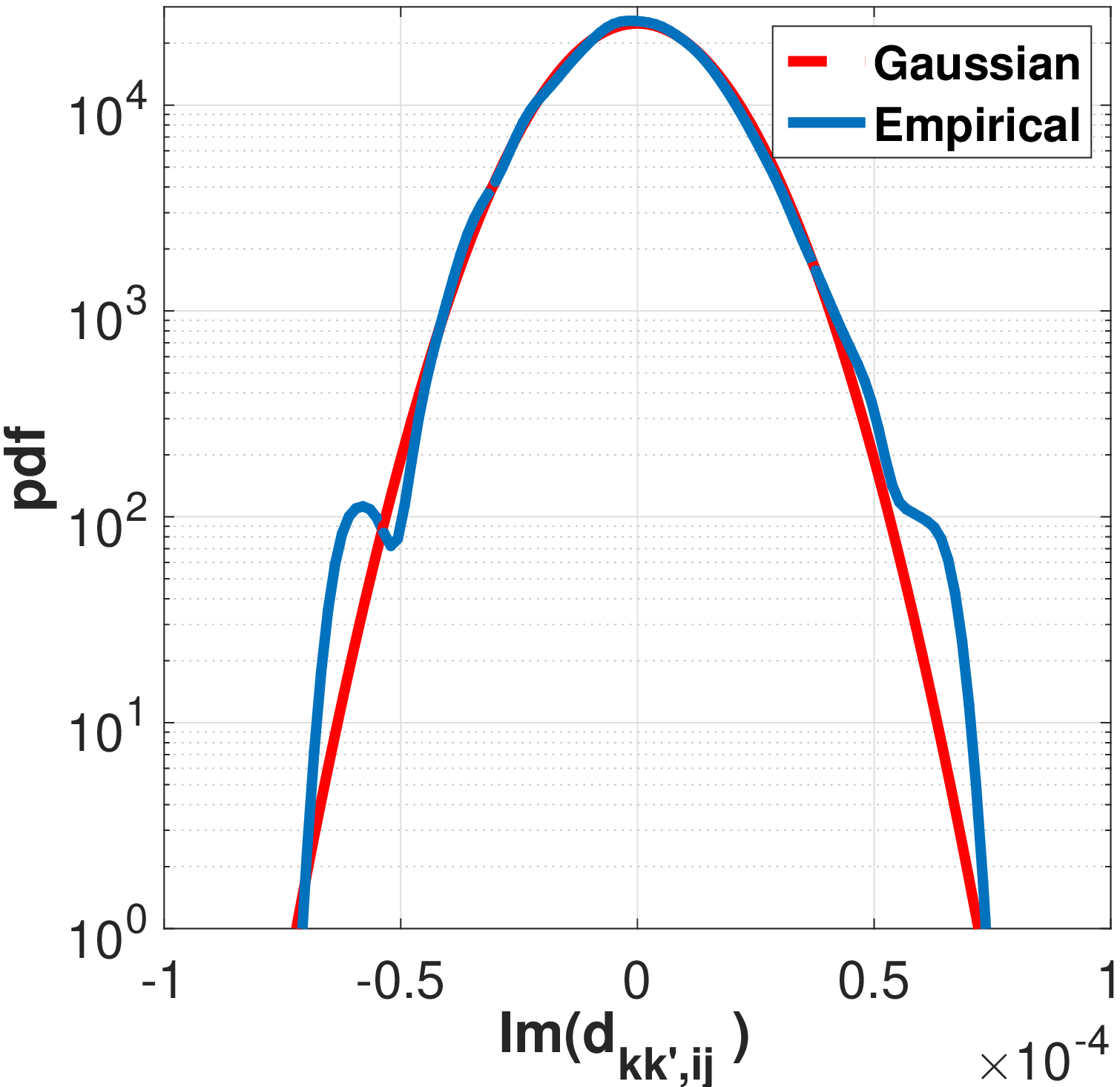}
\caption{True (Empirical) vs approximated (Gaussian) pdf of $d_{kk',ij}), \forall k' \ne k$, and $\forall i,j = 1, 2, \ldots, N$. }
\label{d_kkface}
\end{figure}
Fig. \ref{d_kk} and Fig. \ref{d_kkface} verify the Gaussian approximation in Lemma \ref{lemma3}.  Fig. \ref{d_kk} shows that the probability density functions (pdfs) of the empirical and the Gaussian distribution of $d_{kk,ij}$ ($\forall i,j= 1, 2, \ldots, N$) are very close. Moreover, with the high probability, the imaginary part of $d_{kk,ii}$ is much smaller than the real part, so it can be neglected. In Fig. \ref{d_kkface}, it is clear that the pdfs of both the real part and the imaginary part of $d_{kk',ij}$ ($k' \neq k, \forall i,j= 1, 2, \ldots, N$) are very close to the pdfs of their Gaussian approximations. In both Fig. \ref{d_kk} and Fig. \ref{d_kkface}, apart from the imaginary part of $d_{kk,ii}$ which can be neglected,  the probability of mismatch between approximated and enprirical Gaussian pdfs is very small.

By using Gaussian approximation of $\textbf{D}_{kk}$, we obtain the following approximating closed-form expression of the SE.
\begin{theorem}\label{theo_SE_protocol2_UC}
Given estimated CSI, says $\mathbf{\Theta}_{k} = \{\hat{\textbf{D}}_{k1}, \hat{\textbf{D}}_{k2}, \ldots , \hat{\textbf{D}}_{kK}\} $, and using MMSE-SIC detectors, the achievable downlink SE of the $k$-th user in (\ref{R_k1}) can be rewritten as
\begin{align}\label{I_qrThe2}  
R^{\text{Et-SIC}}_k \!= \!\left(\!1 - \frac{\tau_{\text{u}} \!+\! \tau_{\text{d}}}{\tau_{\text{c}}}\!\right)\! \mathbb {E} \left \{ \log_{2} \left|\textbf{I}_{N} \!+\! \rho\hat{\textbf{D}}^H_{kk} \left(\mathbf{\Psi}^{c'}_{kk}\right)^{-1} \hat{\textbf{D}}_{kk} \right| \right \},
\end{align}
where 
\begin{align}\label{psi_ckk_final}
\mathbf{\Psi}^{c'}_{kk} = \rho\sum^{K}_{k'\neq k}\hat{\textbf{D}}_{kk'}\hat{\textbf{D}}^H_{kk'} + \rho \sum^{K}_{k'=1}\tilde{\textbf{D}}^{\text{var}}_{kk'}  +  \textbf{I}_{N}, 
\end{align}
with
\begin{align}\label{} 
\tilde{\textbf{D}}^{\text{var}}_{kk'} =\colvec{\sum\limits^{N}_{j=1} \text{var}(\tilde{d}_{kk',1j}) & 0  & \hdots & 0  \\
0 & \sum^{N}_{j=1} \text{var}(\tilde{d}_{kk',2j})  & \hdots & 0\\
\vdots
		& & \ddots  &\\
   0 & 0 & \hdots & \sum^{N}_{j=1} \text{var}(\tilde{d}_{kk',Nj}) }, 
\end{align}
where 
\begin{equation}\label{}
\text{var}(\tilde{d}_{kk',ij}) =
\begin{cases}
\frac{\xi_{kk,i} + \kappa_{kk,i}^2}{\tau_{\text{d}}\rho_{\text{d,p}}(\xi_{kk,i} + \kappa_{kk,i}^2) + 1} & \text{if } k = k',\text{and } i = j \\
\frac{\xi_{kk',j}}{\tau_{\text{d}}\rho_{\text{d,p}}\xi_{kk',j} + 1}  & \text{otherwise }.
\end{cases}
\end{equation}
\end{theorem}
\begin{IEEEproof}
	See appendix \ref{Appendix_proftheo_SE_protocol2_UC}.  
\end{IEEEproof}
\subsection{Achievable Downlink SE with Linear MMSE Detectors} 
In this section, we derive the downlink SE using linear MMSE detectors instead of MMSE-SIC detectors to evaluate the difference of the system performance between them. Note that compare with the MMSE-SIC detection, the linear MMSE detection is simpler for the implementation. 
\subsubsection{Protocol 1 with Linear MMSE Detectors} 
At each user, linear MMSE detectors detect $N$ data streams, which are intended for $N$antennas, independently. Then, the linear MMSE detector for the $n$-th data stream of the $k$-th user is given by \cite{li2016massive, ngo2013energy}
\begin{align}
\textbf{f}_{k,n} = \mathbf{\Psi}^{b'}_{kk} \bar{\textbf{d}}_{kk,n},
\end{align}
where $\mathbf{\Psi}^{b'}_{kk} = (\mathbf{\Psi}^b_{kk})^{-1} + \bar{\textbf{D}}_{kk}\bar{\textbf{D}}^H_{kk}$ and $\bar{\textbf{d}}_{kk,n}$ denotes the $n$-th column of $\bar{\textbf{D}}_{kk}$.

Given MMSE detectors at the users, the achievable downlink SE of the $k$-th user can be calculated as
\begin{align}\label{}  
R^{\text{St-MMSE}}_k = \left(1 - \frac{\tau_{\text{u}}}{\tau_c}\right) \sum^{N}_{n=1}\mathbb {E} \left \{ \log_{2} (1 + \zeta^{\text{St-MMSE}}_{k,n}) \right \},
\end{align}
where
\begin{align}\label{}  
\zeta^{\text{St-MMSE}}_{k,n} = \frac{\left|\textbf{f}_{k,n} \bar{\textbf{d}}_{kk,n} \right|^{2}}{\textbf{f}^{H}_{k,n} { \mathbf{\Psi}^{b'}_{kk}  } \textbf{f}_{k,n}   - \left|\textbf{f}_{k,n} \bar{\textbf{d}}_{kk,n} \right|^{2}}.
\end{align}

\subsubsection{Protocol 2 with Linear MMSE Detectors}
Let us denote $\hat{\textbf{d}}_{kk,n}$ as $n$-th column of $\hat{\textbf{D}}_{kk}$. Then, the linear MMSE detector for the $n$-th data stream of the $k$-th user is given by
\begin{align}
\textbf{p}_{k,n} = \mathbf{\Psi}^{c'}_{kk} \hat{\textbf{d}}_{kk,n},
\end{align}
where $\mathbf{\Psi}^{c'}_{kk} = (\mathbf{\Psi}^c_{kk})^{-1} + \hat{\textbf{D}}_{kk}\hat{\textbf{D}}^H_{kk}$.

Given MMSE detectors at the users, the achievable downlink SE of the $k$-th user can be calculated as
\begin{align}\label{I_qrMMSE1}  
R^{\text{Et-MMSE}}_k  =\left(1 - \frac{\tau_{\text{u}} + \tau_{\text{d}}}{\tau_c}\right) \sum^{N}_{n=1}\mathbb {E} \left \{ \log_{2} (1 + \zeta^{\text{Et-MMSE}}_{k,n}) \right \},
\end{align}
where 
\begin{align}\label{}  
\zeta^{\text{Et-MMSE}}_{k,n} = \frac{\left|\textbf{p}_{k,n} \hat{\textbf{d}}_{kk,n} \right|^{2}}{\textbf{p}^{H}_{k,n} { \mathbf{\Psi}^{c'}_{kk}  } \textbf{p}_{k,n}   - \left|\textbf{p}_{k,n} \hat{\textbf{d}}_{kk,n} \right|^{2}}.
\end{align}
In the section \ref{Numsec}, we will compare the performance of the cell-free massive MIMO system when using MMSE-SIC detectors and the one using MMSE detectors at the users. 
\subsection{Achievable Downlink SE given Perfect CSI at the Users}\label{UpperBound}
In this section, we consider a cell-free massive MIMO system with perfect CSI at the users. Although, this is impractical, its performance is considered as the upper bound for the performance of protocol 2. Given perfect CSI, i.e., $\mathbf{\Theta}_{k} = \{\textbf{D}_{k1}, \textbf{D}_{k2}, \ldots , \textbf{D}_{kK}\}$, then the achievable downlink SE of the $k$-th user in (\ref{R_k1}) when using MMSE-SIC detectors at the users is 
\begin{align}\label{I_qrThe3}  
R^{\text{up}}_k = \left(1 - \frac{\tau_{\text{u}} + \tau_{\text{d}}}{\tau_c}\right) \mathbb {E} \left \{ \log_{2} \left| \textbf{I}_{N} + \rho\textbf{D}^H_{kk} \left(\mathbf{\Psi}^d_{kk}\right)^{-1} \textbf{D}_{kk} \right| \right \},
\end{align}
where 
\begin{align}
\mathbf{\Psi}^d_{kk} = \rho\sum^{K}_{k'\neq k}\textbf{D}_{kk'}\textbf{D}^H_{kk'} + \textbf{I}_{N}. \notag
\end{align}

\section{Max-min Power Control}\label{max-min-pc}
In this section, max-min fairness PC is applied for the first protocol to improve the SE of the system. This power control is recomputed on the large-scale fading time scale which changes very slowly. For the second protocol, as the closed-form expression of the SE is very complicated, we apply the power control coefficients of the optimization problem in the first protocol to achieve a sub-optimal solution for the achievable downlink SE of this protocol.
\subsection{Max-min Power Control for Protocol 1}\label{max-min-pc-p1}
In this part, we consider that mutual orthogonal pilot sequences are used in the uplink channel estimation phase. To achieve the fairness good SE for all users in the system, max-min power control is applied for all users to optimize the downlink SE. The max-min fairness optimization problem can be written as
\begin{align}\label{max_min_SE}
&\max _{\{\eta _{mk}\}}~\min \limits _{k=1, \cdots , K} R^{\text{St-SIC}}_k \\
&\text {s.t.}~\sum _{k=1}^{K} \eta _{mk}\gamma _{mk} \leq \frac{1}{LN}, ~ m=1,\ldots , M \notag\\
&\qquad \qquad \quad ~ \eta _{mk} \geq 0, ~ k=1,\ldots , K, ~ m=1,\ldots , M, \notag
\end{align}
where the first constraint of (\ref{max_min_SE}) is the power constraint in (\ref{powcontraint}) when using the mutual orthogonal pilot sequences for the uplink channel estimation phase, and $\gamma _{mk} = \frac{\tau_{\text{u}}\rho_{\text{u}}\beta^2_{mk}}{\tau_{\text{u}}\rho_{\text{u}}\beta_{mk}+ 1}$. 
 An equivalent form of (\ref{max_min_SE}) is
\begin{align*}
&\max _{\{\eta _{mk}\}}~\min \limits _{k=1, \cdots , K} \frac { \left ({\sum _{m=1}^{M} \gamma _{mk}\varsigma _{mk} }\right )^{2} }{\frac{N}{L}\sum \limits _{m=1}^{M}\beta _{mk} \sum \limits _{k'=1}^{K} \gamma _{mk'}\varsigma _{mk'}^{2} +\frac {1}{ \rho L^2} } \\[0.5pt]
&\text {s.t.}~\sum _{k=1}^{K} \eta _{mk}\varsigma^2 _{mk} \leq \frac{1}{LN}, ~ m=1,\ldots , M \\[0.5pt]
&\qquad \eta _{mk} \geq 0, \quad k=1,\ldots , K, ~ m=1,\ldots , M, 
\end{align*}
where $\varsigma _{mk} \triangleq \eta^{1/2} _{mk}$. By introducing the slack variable $\vartheta$, we reformulate (\ref{max_min_SE}) as follows:
\begin{align}\label{max_min_SNR}
&\max _{\{\varsigma _{mk}, \vartheta _{m}\}} ~\min _{k=1, \cdots , K} \frac { \left ({\sum _{m=1}^{M} \gamma _{mk}\varsigma _{mk} }\right )^{2}}{\frac{N}{L} \sum \limits _{m=1}^{M}\!\!\beta _{mk} \vartheta _{m}^{2} +\frac {1}{ \rho L^2}} \\
&\text {s.t.} ~\sum _{k'=1}^{K} \gamma _{mk'}\varsigma _{mk'}^{2} \leq \vartheta _{m}^{2} , ~ m=1,\ldots , M \tag{32a}\\
&\qquad \qquad \quad ~ 0 \leq \vartheta _{m} \leq \frac{1}{\sqrt{LN}}, ~ m=1,\ldots , M \tag{32b}\\
&\qquad \qquad \quad ~ \varsigma _{mk} \geq 0, ~ k=1,\ldots , K, ~ m=1,\ldots , M. \tag{32c} 
\end{align}
By introducing the slack variable $t$, optimization problem (\ref{max_min_SNR}) can be rewritten as 
\begin{align}\label{max_min_SNR_t}
&\max _{\{\varsigma _{mk}, \vartheta _{m}, t\}} t \\ 
&\text {s.t.} ~ t \leq \frac { \left ({\sum _{m=1}^{M} \gamma _{mk}\varsigma _{mk} }\right )^{2}}{\frac{N}{L} \sum \limits _{m=1}^{M}\!\!\beta _{mk} \vartheta _{m}^{2} +\frac {1}{ \rho L^2}}, ~ k=1,\ldots , K,\notag\\
& (32a), (32b), (32c).\notag
\end{align}
Optimization problem (\ref{max_min_SNR_t}) is quasi-concave, as fixing $t$, the problem is second-order cone concave. Therefore, the problem (\ref{max_min_SNR}) is quasi-concave optimization problem and it can be solved effectively by bisection algorithm \cite{boyd2004convex}. 
\subsection{Max-min Power Control for Protocol 2}\label{max-min-pc-sch2}
In this section, similarly to the section \ref{max-min-pc-p1}, we also consider the max-min fairness optimization problem by applying PC. However, it is very difficult (may be impossible) and complicated to obtain the optimal solution for max-min power control of protocol 2, due to the intractable form of the spectral efficiency (\ref{I_qrThe2}). To alleviate such difficulty, we use the power control coefficients from (\ref{max_min_SE}) for protocol 2 which results in sub-optimal performance. 

\section{Numerical results and Discussion}\label{Numsec}
In this section, we provide the numerical results to verify our analytical results and evaluate the performance of cell-free massive MIMO for multiple antennas at both the APs and the users, with and without downlink pilots. Firstly, Gaussian approximations in Lemma \ref{lemma3} are verified and illustrated numerically. Then, we compare the performances of two protocols based on the different CSI available at the users: statistical CSI and estimated CSI. Moreover, the performances of both protocols using MMSE-SIC detectors are also compared with those in systems using MMSE detectors. Finally, the effects of the number of antennas per user, per AP and number of users are analyzed to propose the framework for achieving sub-optimal system performance.

\subsection{Simulation setup}
We assume that the locations of the $M$ APs and $K$ users are uniformly distributed at random within a square of size $1\times1$ km$^2$. Wrapped around technique is used to avoid the boundary effects. In all examples we assume that $\tau_{\text{u}}= \tau_{\text{d}}=K\times N$ and mutually orthogonal pilot sequences are used for both the uplink and the downlink training phase. We also use the same simulation setup with the one in \cite{ngo2017cell}. More specifically, we use i.i.d. Rayleigh fading channels with the three-slope path loss model and shadowing correlation model. The carrier frequency is 1.9 GHz and $\tau_c$ is 300 samples.  

\subsection{Closed-form expression and power control}

\begin{figure}[!h]
\centering
\includegraphics[width=0.49\textwidth]{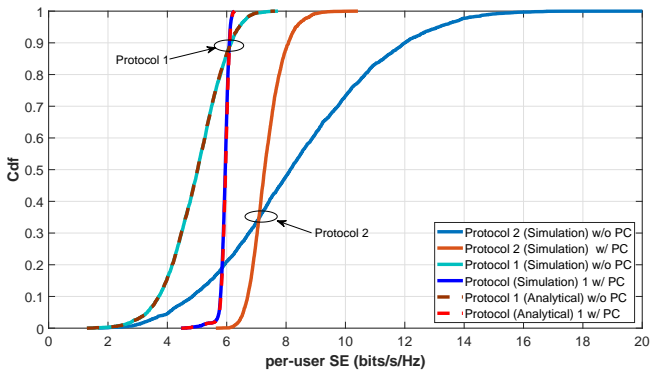}
\caption{Cummulative distribution function (cdf) of per-user SE with $M$ = 50, $K$ = 10, $N$ = 2, $L$ = 4.}
\label{closed_form}
\end{figure}

\begin{figure}[!h]
\centering
\includegraphics[width=0.49\textwidth]{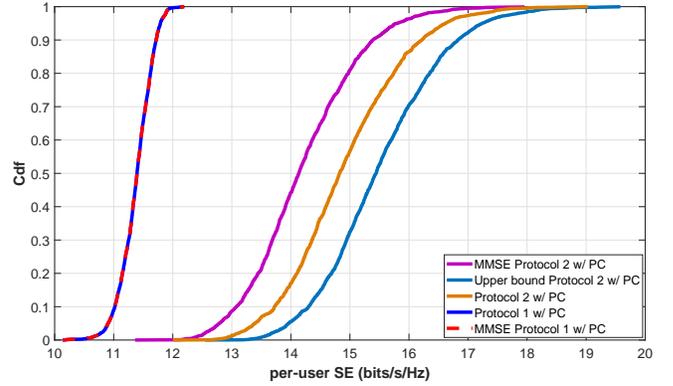}
\caption{Cdf of per-user SE with $M$ = 50, $K$ = 5, $N$ = 4, $L$ = 4.}
\label{fig_full}
\end{figure}

\begin{figure}[!h]
	\centering
	\includegraphics[width=0.49\textwidth]{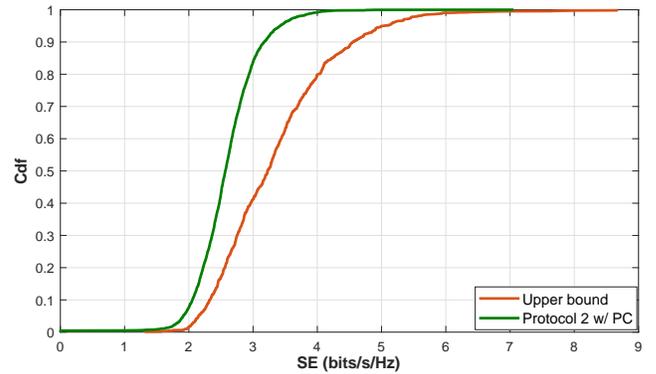}
	\caption{Cdf of per-user SE with $M$ = 20, $K$ = 5, $N$ = 1, $L$ = 1.}
	\label{fig_uppbound}
\end{figure}
Fig. \ref{closed_form} numerically proves the correction of our derived closed-form expression in Theorem \ref{theo_closedform}. For transmission protocol 1, this figure shows the perfect match between the SEs, which are derived analytically (closed-form expression), and the SEs, which are conducted by simulation of (\ref{R_k_statistic}), regardless of using PC or not.
In this paper, max-min fairness power control is applied to provide the good uniform service to all users in the systems. By using closed-form of the downlink SE of protocol 1 given statistical CSI at the users, the power control of the downlink SE can be solved effectively by bisection algorithm. Due to the computation complexity of SE optimization problems in the second protocol, we apply the power control coefficients of the first protocol to the second protocol. Interestingly, Fig. \ref{closed_form} shows that, in term of 95\% likely per-user SE, not only the SE of  protocol 1 increases dramatically, by about 80\% with PC, but also the SE of protocol 2 improves significantly, by about 60\%. This verifies that the optimal PC coefficients of protocol 1 can be applied effectively to PC problem of protocol 2. 

Figure \ref{fig_full} shows that the performance of protocol 2 is very close to its upper bound, when perfect channel
information is available at the users. This implies that our PC method applies
effectively on protocol 2. Moreover, this figure also compares the performance of two protocols using different detection techniques at the users, i.e., MMSE-SIC and MMSE. In protocol 1, there is no difference in SE between the system using MMSE-SIC detectors and the one using MMSE detectors. This completely agrees with the conclusion in \cite{li2016massive}, as only statistical CSI is available at the users. However, in protocol 2, as SIC works effectively with estimated CSI at the users, the SE of the system using MMSE-SIC increases noticeably compared to the one using MMSE. This gap varies depending on the number of users, $K$, and the number of antennas per user, $N$, which is shown in Fig. \ref{N_change}. Specially, when $N = 1$, the performances of this protocol are exactly identical regardless either using MMSE-SIC or MMSE detectors.

To further see the performance limit as well as how well our sub-optimal power control in Section~\ref{max-min-pc-sch2}  is, we compare the SE of protocol 2 using our sub-optimal power control with the SE of the ideal case where users have perfect CSI and optimal power control is performed. For simplicity, we consider a special case where each user/AP is equipped with a  single-antenna, and mutual orthogonal pilot sequences are used in both the uplink and downlink channel estimation phases. From (\ref{I_qrThe3}), and by using the approximation   $\mathbb {E} \left \{\log_2(1+X/Y)\right\} \approx \log_2\left(1 + \mathbb {E} \left \{X\right\}/\mathbb {E} \left \{Y\right\} \right)$ \cite{zhang2014power}, the spectral efficiency with perfect CSI at user $k$ can be approximated by
\begin{align}\label{eq_up3}  
&R^{\text{up-approx}}_k =  \left(1 - \frac{\tau_{\text{u}} + \tau_{\text{d}}}{\tau_c}\right)\times\notag\\
&\quad \times \log_{2} \left(1 + \frac { \left ({\sum _{m=1}^{M} \gamma _{mk}\varsigma _{mk} }\right )^{2} + \sum \limits _{m=1}^{M}\beta _{mk} \gamma _{mk}\varsigma _{mk}^{2}}{\sum \limits _{m=1}^{M}\beta _{mk} \sum \limits _{k'\ne k}^{K} \gamma _{mk'}\varsigma _{mk'}^{2} +\frac {1}{ \rho} }\right).
\end{align}
Therefore, the corresponding max-min power control can be efficiently solved  by using the
successive approximation technique \cite{tran2012fast}. 

Figure~\ref{fig_uppbound} shows the per-user SE of protocol 2 with sub-optimal power control in section IV-B, and the one with perfect CSI at the user and optimal power control. We can see that the performance gap is quite small. This verifies that the  sub-optimal power control works very well.
\subsection{Effects of the number of users, number of antennas per APs and per users}
\begin{figure}[!h]
\centering
\includegraphics[width=0.49\textwidth]{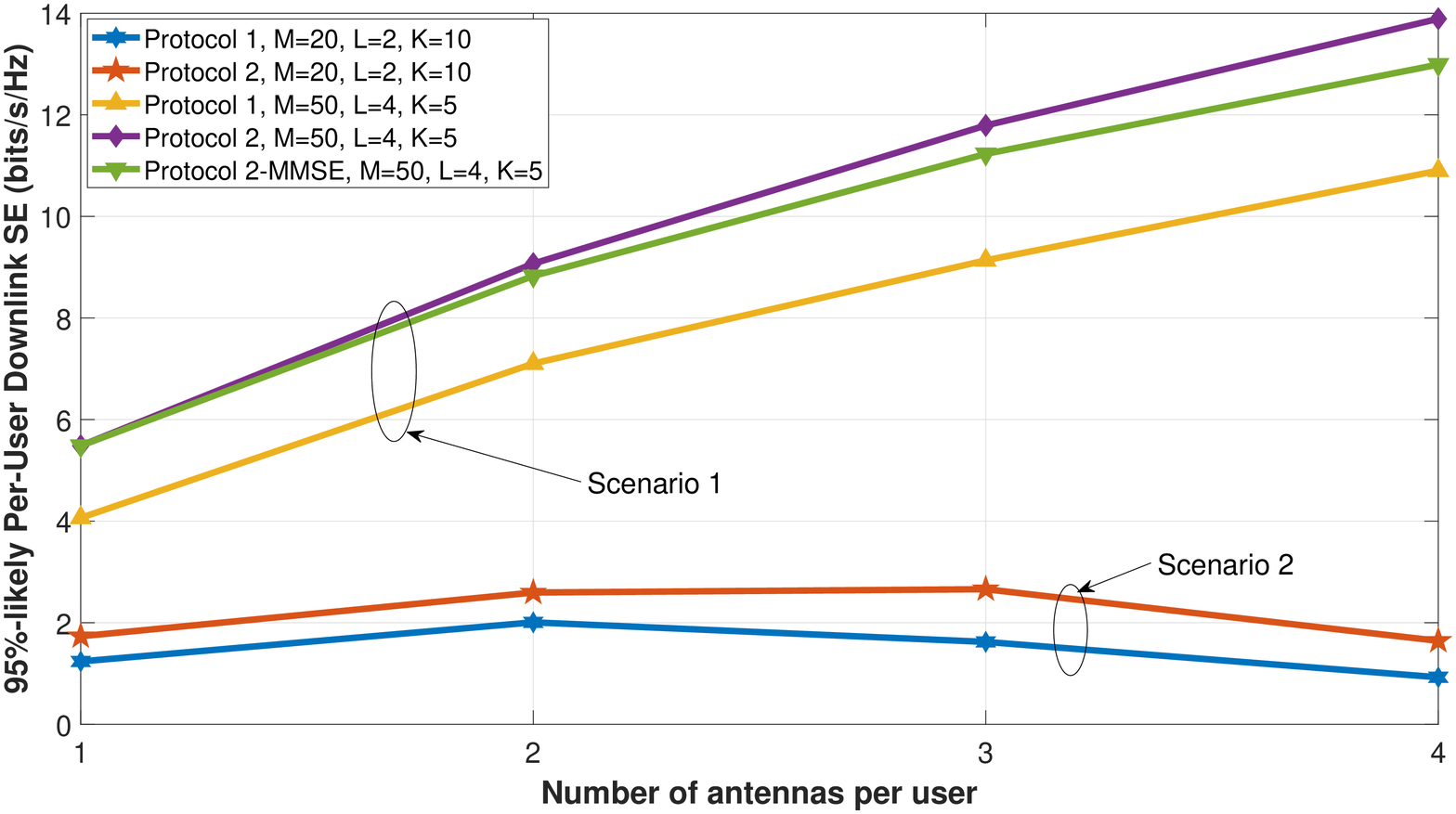}
\caption{95$\%$-likely per-user downlink SE v.s. number of antennas per user.}
\label{N_change}
\end{figure}

\begin{figure}[!h]
	\centering
	\includegraphics[width=0.49\textwidth]{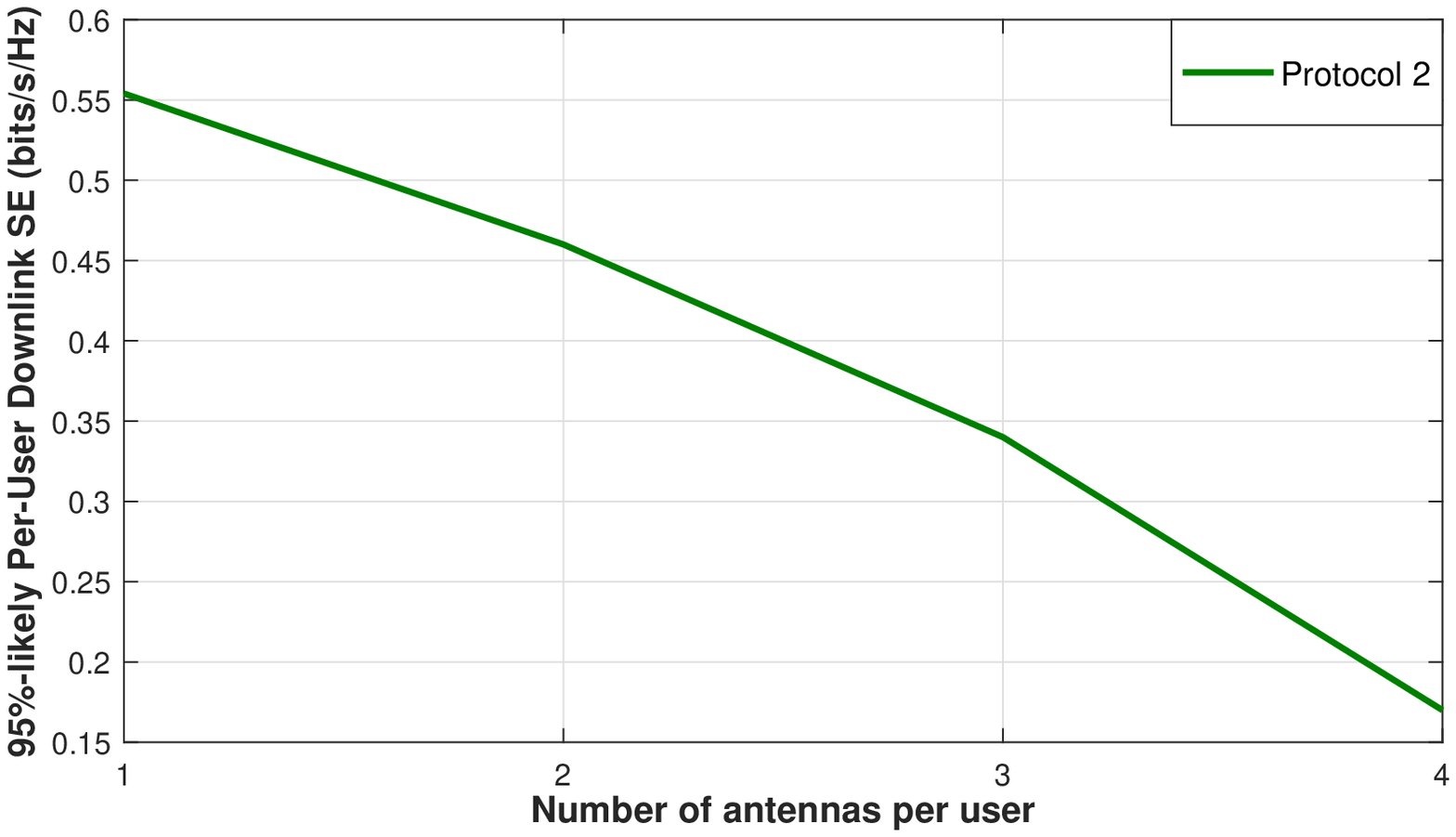}
	\caption{95$\%$-likely per-user downlink SE v.s. number of antennas per user  with $M$ = 20, $K$ = 30, $L$ = 1.}
	\label{N_change_p2}
\end{figure}

\begin{figure}[!h]
\centering
\includegraphics[width=0.49\textwidth]{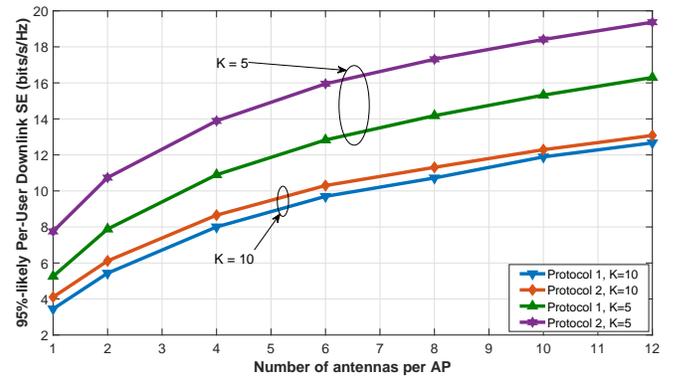}
\caption{95$\%$-likely per-user downlink SE with $M$ = 50, $L$ = 4, $N$ = 4.}
\label{L_change}
\end{figure}

\begin{figure}[!h]
\centering
\includegraphics[width=0.49\textwidth]{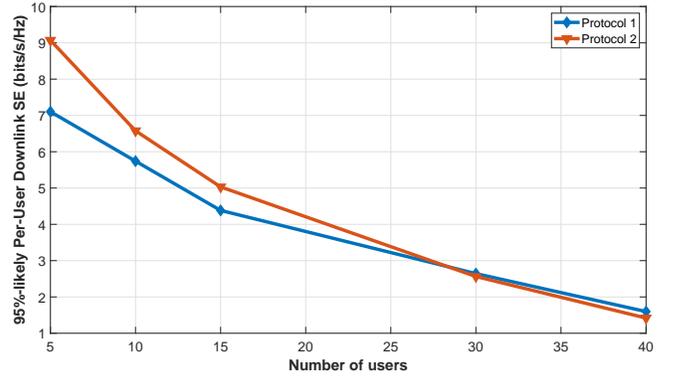}
\caption{95$\%$-likely per-user downlink SE v.s number of users. Here, $M$ = 50, $L$ = 4, $N$ = 2.}
\label{K_change}
\end{figure}
First, in Fig. \ref{N_change}, the effect of number of antennas per user is investigated by using two scenarios with different number of APs and number of antennas per APs.  In scenario 1, with small number of users in comparison with number of APs and number of total antennas at the APs, the per-user downlink SE increases proportionally with the number of antennas per user as the increasing of independent channels (or degrees of freedom) per user. Using multiple antennas at the users greatly enhances the per-user SE, especially with small number of users in the system. With $K=5$, by using 3 antennas per user and PC, we can double the 95$\%$-likely SE of both protocols, compared to single-antenna user systems. However, in scenario 2, the per-user downlink SE first increases when the number of antennas per user increases with increasing independent channels (or degrees of freedom) per user. Then, this SE will reach the maximum value and then it will decrease when the number of antennas per user increases. Especially, when the number of users is large,  single-antenna user setups  outperform multi-antenna user setups. The reason is that although the number of independent channels per user increases, the channel estimation overhead  also increases. This channel estimation overhead largely dominates when $N$ is large.

Next, we evaluate the effect of AP antennas. In Fig. \ref{L_change}, 95$\%$-likely per-user achievable downlink SE of two protocols with PC are shown with arbitrary number of antennas per APs. As expected, the performances of all protocols increase proportionally with the number of antennas per AP, especially when the total number of antennas at the users is small in comparison with the total number of antennas at the APs (here, $M \times L = 200$). This comes from the fact that when $L$ increases, the channel is more favorable, and hence, the inter-user interference reduces. At the same time, the array gain increases. 

Finally, 95$\%$-likely per-user achievable downlink SE of two protocols are shown in Fig. \ref{K_change} with arbitrary number of users and $N = 2$. At first, when the number of user is small, protocol 2 achieves higher SE than protocol 1. However, when the number of users increases, protocol 1 achieves higher SE than protocol 2, as a large proportion of coherence duration is assigned for the downlink training. These above insights are very important for us to design the framework for achieving the sub-optimal SE of the system, in the next section.    

\subsection{Framework for achieving the sub-optimal SE}
Normally, the number of APs and number of antennas per APs are fixed. However, the number of users in the system, which strongly affects on the system performance, is unknown. Therefore, we propose the framework that automatically chooses suitable protocol as well as the number of active antennas per users, based on the number of users in the systems. The framework is as follows

\begin{itemize}
\item In the setup session, based on the real number of users, $K$, and their locations, the system can calculate its SE for each protocol and for each number of active antennas per user, $n$, $\forall n = 1, 2, \hdots, N$.
\item Choose the protocol with the number of active antennas per user, $n$, which achieves the highest SE. 
\end{itemize}
The framework only need to be updated with the time frame of PC, i.e., updated infrequently, as it just depends on the large-scale fading and number of active users.
\begin{remark}
The SE, which is a result of the framework, is sub-optimal, as it is considered under following conditions: using mutual orthogonal pilot sequences, using sub-optimal power control for protocol 2, and using the same number of active antennas per users. 
\end{remark}
\section{Conclusion}\label{Concl}
In this paper, we evaluated the downlink SE of two transmission protocols of cell-free massive MIMO system with finite numbers of APs, users and arbitrary numbers of antennas at the APs and the users. There are no downlink pilots in the first transmission protocol, while the downlink pilots were beamformed to the used for the CSI acquisition in the second transmission protocol. Compared with the first protocol, the second protocol has higher channel estimation quality at the users, but higher channel estimation overhead. Numerical results show that no protocol always shows the advantage over the other in the performance as they both depend on the number of users and number of antennas per user in the system. Finally, by using two protocols and combining with multi-antenna users, the system can achieve the sub-optimal performance regardless of number of users in the system.

\appendix
\subsection{Proof of Lemma \ref{lemma1}}\label{AppendixA}
From (\ref{vecchann}), we have
\begin{align} 
\text{vec}(\hat{\textbf{G}}_{mk}) &=\sqrt{\tau_{\text{u}}\rho_{\text{u}}}\beta_{mk}\left(\tau_{\text{u}}\rho_{\text{u}}\sum_{i=1}^{K}\beta_{mi}\pmb{\Phi}^T_{\text{u},ik}\pmb{\Phi}^*_{\text{u},ik}+ \textbf{I}_{N}  \right)^{-1}\notag\\
&\quad\otimes\textbf{I}_{L}\text{vec}(\textbf{Y}_{\text{u},mk}). \label{eqvec}
\end{align}
Finally, (\ref{estchan1}) is obtained by applying the following identity $\text{vec}(ABC)=C^T \otimes A \text{vec}(B)$ on (\ref{eqvec}).

\subsection{Proof of Lemma \ref{lemma2}}\label{AppendixB}
We denote $i$-th column of $\textbf{G}_{mk}$ is $\textbf{g}_{mk,i}$, $j$-th column of $\hat{\textbf{G}}_{mk}$ is $\hat{\textbf{g}}_{mk,j}$, and $\textbf{G}_{mk}$ can be written as $\textbf{G}_{mk} = \colvec{\textbf{g}_{mk,1} & \textbf{g}_{mk,2}  & \hdots & \textbf{g}_{mk,N}}$. Then the downlink matrix channel of $k$-th user can be written as
\begin{align} 
\textbf{D}_{kk} = \colvec{d_{kk,11} & d_{kk,12} & \hdots & d_{kk,1N} \\
d_{kk,21} & d_{kk,22} & \hdots & d_{kk,2N}\\
\vdots
		& & \ddots  &\\
   d_{kk,N1} &  &  & d_{kk,NN}},
\end{align}
where $d_{kk,ij} \triangleq \sum_{m=1}^{M}\eta_{mk}^{1/2}\textbf{g}^H_{mk, i}\hat{\textbf{g}}_{mk, j}$. To estimate the element $d_{kk,ij}$ of the matrix channel $\textbf{D}^{kk}$, firstly $\textbf{y}^T_{\text{dp},ki}$ is projected onto $\phi_{\text{d},kj}$ to obtain
 \begin{align} 
y_{\text{dp},k,ij} = \textbf{y}^T_{\text{dp},k,i}\phi_{\text{d},k,j}.
\end{align}
Then, MMSE estimation of $d_{kk,ij}$ given $y_{\text{dp},k,ij}$ is calculated as follows
\begin{align}\label{mmse_a_dl} 
\hat{d}_{kk,ij} &= \mathbb {E}\{d_{kk,ij}\}+ \textbf{C}_{d_{kk,ij},y_{\text{dp},k,ij}}\times \notag\\
&\quad\times\textbf{C}^{-1}_{y_{\text{dp},k,ij},y_{\text{dp},k,ij}}\left(y_{\text{dp},k,ij} - \mathbb {E} \{ y_{\text{dp},k,ij} \}  \right),
\end{align}
where $\textbf{C}_{x,y}$ denotes the covariance of $x$ and $y$.

\subsubsection{Compute $\hat{d}_{kk,ij}$ with $i = j$}\label{AppendixB1}
\begin{align}\label{Ea2}
\mathbb {E}\{d_{kk,ii}\} &= \mathbb {E}\left\{ \sum_{m=1}^{M}\eta_{mk}^{1/2}\textbf{g}^H_{mk, i}\hat{\textbf{g}}_{mk, i}\right\}\notag\\
&= L\sum_{m=1}^{M}\eta_{mk}^{1/2}\gamma_{mk,i},
\end{align}
where $\gamma_{mk,i} \triangleq \left[\mathbb {E}\left\{ \hat{\textbf{g}}_{mk, i}\hat{\textbf{g}}^H_{mk, i}\right\} \right ]_{nn}, \forall n \in N$. Note that $\gamma_{mk,i}$ not depends on the index of antenna in the $m$-th AP. Then, from (\ref{vecchann} ), $\gamma_{mk,i}$ can be calculated as
\begin{align}
&\gamma_{mk,i}\notag\\ 
&= \left[\mathbb {E}\left\{ \text{vec}(\hat{\textbf{G}}_{mk})\text{vec}(\hat{\textbf{G}}_{mk})^H  \right\} \right ]_{(i-1)L + l}, \forall l \in L\notag\\ 
&= \left[\! \tau_{\text{u}}\rho_{\text{u}}\beta^2_{mk} \left(\tau_{\text{u}}\rho_{\text{u}}\sum_{i=1}^{K}\tilde{\pmb{\Phi}}_{\text{u},ik}\beta_{mi}\tilde{\pmb{\Phi}}^H_{\text{u},ik}+ \textbf{I}_{LN}  \right)^{-1}\! \right ]_{(i-1)L + l}.
\end{align}
Furthermore, we have
\begin{align}
\mathbb {E} \{ y_{\text{dp},k,ii} \}\label{Ey2}  &=  \mathbb {E}\left\{    \sqrt{\tau_{\text{d}}\rho_{\text{d,p}}} \sum_{m=1}^{M}\eta_{mk}^{1/2}\textbf{g}^H_{mk, i}\hat{\textbf{g}}_{mk, i} + w_{k,i} \right\}\notag\\
&= L\sqrt{\tau_{\text{d}}\rho_{\text{d,p}}}\sum_{m=1}^{M}\eta_{mk}^{1/2}\gamma_{mk,i},\\
\textbf{C}_{d_{kk,ii},y_{\text{dp},k,ii}}\label{Cay2}  &= \mathbb {E} \Biggl\{ \left(\sum_{m=1}^{M}\eta_{mk}^{1/2}\textbf{g}^H_{mk, i}\hat{\textbf{g}}_{mk, i}\right)\times\notag\\
&\times\left( \sqrt{\tau_{\text{d}}\rho_{\text{d,p}}} \sum_{m=1}^{M}\eta_{mk}^{1/2}\textbf{g}^H_{mk, i}\hat{\textbf{g}}_{mk, i} + w_{k,i}   \right) \Biggl\} \notag\\
&= \sqrt{\tau_{\text{d}}\rho_{\text{d,p}}}\mathbb {E} \Biggl\{ \left(\sum_{m=1}^{M}\eta_{mk}^{1/2}\textbf{g}^H_{mk, i}\hat{\textbf{g}}_{mk, i}\right)^2\Biggl\} \notag\\
&= \sqrt{\tau_{\text{d}}\rho_{\text{d,p}}}(P_1 + P_2),
\end{align}
where $s_{mk} \triangleq \eta_{mk}^{1/2}\textbf{g}^H_{mk, i}\hat{\textbf{g}}_{mk, i}$, $P_1 \triangleq \mathbb {E} \Bigl\{ \sum_{m=1}^{M}s_{mk}^2 \Bigl\}$ and $P_2 \triangleq \mathbb {E} \Bigl\{\sum_{m=1}^{M}\sum_{n \neq m}^{M}s_{mk}s_{nk} \Bigl\}$.

Compute $P_1$:
\begin{align}\label{P1}  
P_1 &= \sum_{m=1}^{M}\mathbb {E} \Bigl\{s_{mk}^2 \Bigl\} \notag\\
&= \sum_{m=1}^{M}\mathbb {E} \Bigl\{ \Bigl(\eta^{1/2}_{mk}\sum_{l=1}^{L}g_{mk,il}\hat{g}_{mk,il} \Bigl)^2   \Bigl\} \notag\\
&= \eta_{mk}\sum_{m=1}^{M}\Biggl(\mathbb {E} \Bigl\{ \sum_{l=1}^{L}(\hat{g}^4_{mk,il} + \tilde{g}^2_{mk,il}\hat{g}^2_{mk,il})\Bigl\} \notag\\
&\quad+\mathbb {E} \Bigl\{ \sum_{l=1}^{L}\sum_{l'\neq l}^{L} g_{mk,il}\hat{g}_{mk,il} g_{mk,il'}\hat{g}_{mk,il'}   \Bigl\} \Biggl) \notag\\
&\mathop = \limits ^{(a)} \eta_{mk}\sum_{m=1}^{M}(L\beta_{mk}\gamma_{mk,i} + L^2\gamma^2_{mk,i}  ),
\end{align}
where $(a)$ follows the fact that $g_{mk,il}\hat{g}_{mk,il}$ and $g_{mk,il'}\hat{g}_{mk,il'}$ are independent with $\forall l'\neq l$.

Compute $P_{2}$:
\begin{align}\label{P2}  
P_2 = L^2 \sum_{m=1}^{M}\sum_{n \neq m}^{M}\eta_{mk}^{1/2}\eta_{nk}^{1/2} \gamma_{mk,i} \gamma_{nk,i}.
\end{align}
Substituting (\ref{P1}) and (\ref{P2}) into (\ref{Cay2}), we have
\begin{align}  
\textbf{C}_{d_{kk,ii},y_{\text{dp},k,ii}} &=\sqrt{\tau_{\text{d}}\rho_{\text{d,p}}}\Bigg(L\sum_{m=1}^{M}\eta_{mk}\beta_{mk}\gamma_{mk,i} +\notag\\
&\quad+\left(L\sum_{m=1}^{M}\eta^{1/2}_{mk}\gamma_{mk,i} \right)^2 \Bigg).
\end{align}

Next, we have
\begin{align}
\textbf{C}_{y_{\text{dp},k,ii},y_{\text{dp},k,ii}} &= \mathbb {E} \Biggl\{\! \left(\! \sqrt{\tau_{\text{d}}\rho_{\text{d,p}}} \sum_{m=1}^{M}\eta_{mk}^{1/2}\textbf{g}^H_{mk, i}\hat{\textbf{g}}_{mk, i} + w_{k,i} \!  \right)^2 \!\!\Biggl\} \notag\\
&= \mathbb {E} \Biggl\{\! \tau_{\text{d}}\rho_{\text{d,p}}\left(\sum_{m=1}^{M}\eta_{mk}^{1/2}\textbf{g}^H_{mk, i}\hat{\textbf{g}}_{mk, i}\right)^2 + w^2_{k,i}\! \Biggl\}.
\end{align}
Then, following the similar method on computing $\textbf{C}_{d_{kk,ii},y_{\text{dp},k,ii}}$, we have

\begin{align}\label{Cyy2} 
\textbf{C}_{y_{\text{dp},k,ii},y_{\text{dp},k,ii}} = \tau_{\text{d}}\rho_{\text{d,p}}\left(\xi_{kk,i} + \left(L\sum_{m=1}^{M}\eta^{1/2}_{mk}\gamma_{mk,i}\! \right)^2 \right) + 1.
\end{align}
Finally, (\ref{seq}) is derived by the substitution of (\ref{Eay}), (\ref{Cay}) and (\ref{Cyy}) into (\ref{mmse_a_dl}).
\subsubsection{Compute $\hat{d}_{kk,ij}$ with $i \neq j$}\label{AppendixB2}
\begin{align}
\mathbb {E}\{d_{kk,ij}\}\label{Eay}  &= \mathbb {E} \{ y_{\text{dp},k,ij} \} = 0,\\
\textbf{C}_{d_{kk,ij},y_{\text{dp},k,ij}}\label{Cay}  &= \mathbb {E} \Biggl\{ \left(\sum_{m=1}^{M}\eta_{mk}^{1/2}\textbf{g}^H_{mk, i}\hat{\textbf{g}}_{mk, j}\right)\times\notag\\
&\quad\times\left( \sqrt{\tau_{\text{d}}\rho_{\text{d,p}}} \sum_{m=1}^{M}\eta_{mk}^{1/2}\textbf{g}^H_{mk, i}\hat{\textbf{g}}_{mk, j} + w_{k,i}   \right) \Biggl\} \notag\\
&\mathop = \limits ^{(a1)}  L\sqrt{\tau_{\text{d}}\rho_{\text{d,p}}}\sum_{m=1}^{M}\eta_{mk}\beta_{mk}\gamma_{mk,j},\\
\textbf{C}_{y_{\text{dp},k,ij},y_{\text{dp},k,ij}}\label{Cyy}  &= \mathbb {E} \Biggl\{  \left( \sqrt{\tau_{\text{d}}\rho_{\text{d,p}}} \sum_{m=1}^{M}\eta_{mk}^{1/2}\textbf{g}^H_{mk, i}\hat{\textbf{g}}_{mk, j} + w_{k,i}   \right)^2 \Biggl\} \notag\\
&\mathop = \limits ^{(a2)}  L\tau_{\text{d}}\rho_{\text{d,p}}\sum_{m=1}^{M}\eta_{mk}\beta_{mk}\gamma_{mk,j} + 1,
\end{align}
where $(a1)$ and $(a2)$ follow the similar method on computing  $\textbf{C}_{d_{kk,ii},y_{\text{dp},k,ii}}$.
The substitution of (\ref{Eay}), (\ref{Cay}) and (\ref{Cyy}) into (\ref{mmse_a_dl}) yields (\ref{seq}). 
\subsubsection{Compute $\hat{d}_{kk',ij}$ with $k \neq k'$ and $\forall i,j$ } \label{AppendixB3}
The computation is similar to that in Section~\ref{AppendixB2}.

\subsection{Proof of Theorem \ref{theo_SE}}\label{Appendix_theo_SE}
Mutual information is defined as \cite{cover2012elements}
\begin{align}\label{I_qrThe1}  
I(\textbf{q}_{k}; \textbf{r}_{k},\mathbf{\Theta}_{k}) = h(\textbf{q}_{k}|\mathbf{\Theta}_{k}) - h(\textbf{q}_{k}|\textbf{r}_{k},\mathbf{\Theta}_{k}),
\end{align}
where $h(.)$ is the differential entropy, and $\mathbf{\Theta}_{k}$ is the channel information at $k$-th user. Choose suboptimal $\textbf{q}_{k}$ as $\mathcal {CN}\left ({{0},{\textbf{I}_{N}}}\right )$, then
\begin{align}\label{h_qk1} 
h(\textbf{q}_{k}|\mathbf{\Theta}_{k}) = \log_{2}|\pi e \textbf{I}_{N}|.
\end{align}
MMSE estimation of $\textbf{q}_{k}$ in (\ref{data_recei_eq}) given $\textbf{r}_{k}$ and $\mathbf{\Theta}_{k}$  is 
\begin{align} 
\hat{\textbf{q}}_{k} &= \mathbb {E}\{\textbf{q}_{k}|\mathbf{\Theta}_{k}\}+ \sqrt{\rho}\mathbb {E}\{\textbf{D}^H_{kk}|\mathbf{\Theta}_{k}\}\mathbf{\Psi}_{kk} \left(\textbf{r}_{k} - \mathbb {E} \{ \textbf{r}_{k}|\mathbf{\Theta}_{k} \}  \right)\notag\\
&=\sqrt{\rho}\mathbb {E}\{\textbf{D}^H_{kk}|\mathbf{\Theta}_{k}\} \left(\mathbf{\Psi}_{kk}\right)^{-1}  \textbf{r}_{k}, 
\end{align}
where $\mathbf{\Psi}_{kk} = \mathbb {E}\{\rho\sum^{K}_{k'=1} \textbf{D}_{kk'}\textbf{D}^H_{kk'}|\mathbf{\Theta}_{k}\} + \textbf{I}_N.$
Let $\tilde{\textbf{q}}_{k} \triangleq \textbf{q}_{k} - \hat{\textbf{q}}_{k}$ denote the estimation error of $\textbf{q}_{k}$, then following \cite[Appendix I]{yoo2006capacity}, $h(\textbf{q}_{k}|\textbf{r}_{k},\mathbf{\Theta}_{k})$ is upper bounded by 
\begin{align}\label{h_qk_rD1} 
h(\textbf{q}_{k}|\textbf{r}_{k},\mathbf{\Theta}_{k}) &\le \mathbb {E}\left\{\log_{2}\left|\pi e \mathbb {E}\{\tilde{\textbf{q}}_{k} \tilde{\textbf{q}}^H_{k}|\mathbf{\Theta}_{k}\}\right|\right\} \notag\\ 
&= \mathbb {E}\left\{\log_{2} \left| \pi e \left(\textbf{I}_{N} - \mathbf{\Upsilon}_{kk} \right)\right|\right\},
\end{align}
where $\mathbf{\Upsilon}_{kk} = \rho\mathbb {E}\{\textbf{D}^H_{kk}|\mathbf{\Theta}_{k}\} \left(\mathbf{\Psi}_{kk}\right)^{-1}  \mathbb {E}\{\textbf{D}_{kk}|\mathbf{\Theta}_{k}\}$.

Substituting (\ref{h_qk1}) and (\ref{h_qk_rD1}) into (\ref{I_qrThe1}) and applying the matrix inversion lemma, we have
\begin{align}\label{I_qrThe4}  
I(\textbf{q}_{k}; \textbf{r}_{k},\mathbf{\Theta}_{k}) \ge \mathbb {E}\left\{ \log_{2} \left| \textbf{I}_{N} + \mathbf{\Upsilon}^a_{kk} \right|\right\},
\end{align}
where $\mathbf{\Upsilon}^a_{kk} = \rho \mathbb {E}\{\textbf{D}^H_{kk}|\mathbf{\Theta}_{k}\} \left(\mathbf{\Psi}^a_{kk}\right)^{-1}  \mathbb {E}\{\textbf{D}_{kk}|\mathbf{\Theta}_{k}\}$,
and $\mathbf{\Psi}^a_{kk} = \textbf{I}_{N} + \mathbb {E}\{(\rho\sum^{K}_{k'=1}\textbf{D}_{kk'}\textbf{D}^H_{kk'}|\mathbf{\Theta}_{k})\} - \rho \mathbb {E}\{\textbf{D}_{kk}|\mathbf{\Theta}_{k}\} \mathbb {E}\{\textbf{D}^H_{kk}|\mathbf{\Theta}_{k}\}.$
Note that the dimension of invertible matrix $\mathbf{\Psi}^a_{kk}$ only depends on the number of antennas at the users.
Then achievable downlink SE of the $k$-th user when using MMSE-SIC detectors at the receivers can be calculated as
\begin{align} 
R_k = (1 - \tau_{tot}/\tau_c)\mathbb {E}\left\{ \log_{2} \left| \textbf{I}_{N} + \mathbf{\Upsilon}^a_{kk} \right|\right\},
\end{align}
where $\tau_{tot}$ is total training duration per coherence interval $\tau_c$.
\subsection{Lemma \ref{lemma4}}\label{AppendixD}
This Lemma will be used to proof Theorem \ref{theo_closedform}.
\begin{lemma}\label{lemma4} 
Let $\mathbf{B} = \mathbf{Y}^H\mathbf{X}$, where ${\mathbf {X} }$, ${\mathbf {Y} }$ are $M \times N$ random matrix which its elements are assumed to be i.i.d. $\mathcal {CN}\left ({{0},{1}}\right )$ and $\mathbf {C}$ is $N \times N$ matrix. Then
\begin{align} 
&\mathop {\mathrm {\mathbb {E}}}\nolimits \left \{{{{\mathbf {B}^H} {\mathbf {C}} {\mathbf {B} }}}\right \}\notag\\
&= \colvec{\Tr\left(\mathbf {C} \mathop {\mathrm {\mathbb {E}}}\nolimits \left \{\mathbf{b}_1 \mathbf{b}^H_1 \right\}\right) &   &  & \\
   & \Tr\left(\mathbf {C} \mathop {\mathrm {\mathbb {E}}}\nolimits \left \{\mathbf{b}_2 \mathbf{b}^H_2 \right\}\right) &  & \\
   &  & \ddots &  \\
   &  &  & \Tr\left(\mathbf {C} \mathop {\mathrm {\mathbb {E}}}\nolimits \left \{\mathbf{b}_N \mathbf{b}^H_N \right\}\right)},
\end{align}
where
\begin{align}
\mathop {\mathrm {\mathbb {E}}}\nolimits \left \{\mathbf{b}_k \mathbf{b}^H_k \right\} =& \colvec{
  \mathop {\mathrm {\mathbb {E}}}\nolimits \left \{\vert \mathbf{y}^H_1 \mathbf{x}_k \vert^2 \right\} &   &  & \\
   & \mathop {\mathrm {\mathbb {E}}}\nolimits \left \{\vert \mathbf{y}^H_2 \mathbf{x}_k \vert^2 \right\} &  & \\
   &  & \hdots &  \\
   &  &  & \mathop {\mathrm {\mathbb {E}}}\nolimits \left \{\vert \mathbf{y}^H_N \mathbf{x}_k \vert^2 \right\}}.
\end{align}
\end{lemma}
\textit{Proof}:
\begin{align} 
&\mathop {\mathrm {\mathbb {E}}}\nolimits \left \{{{{\mathbf {B}^H} {\mathbf {C}} {\mathbf {B} }}}\right \}\notag\\
&= \mathop {\mathrm {\mathbb {E}}}\nolimits \left \{
 \begin{bmatrix}
  \mathbf{b}^H_1 \\
  \mathbf{b}^H_2\\
  \vdots\\
  \mathbf{b}^H_N
 \end{bmatrix} \mathbf {C}
 \begin{bmatrix}
  \mathbf{b}_1 & \mathbf{b}_2 & \hdots & \mathbf{b}_N
 \end{bmatrix}\right \} \notag \\
&\quad=\colvec{\Tr\left(\mathbf {C} \mathop {\mathrm {\mathbb {E}}}\nolimits \left \{\mathbf{b}_1 \mathbf{b}^H_1 \right\}\right) &   &  & \\
   & \Tr\left(\mathbf {C} \mathop {\mathrm {\mathbb {E}}}\nolimits \left \{\mathbf{b}_2 \mathbf{b}^H_2 \right\}\right) &  & \\
   &  & \ddots &  \\
   &  &  & \Tr\left(\mathbf {C} \mathop {\mathrm {\mathbb {E}}}\nolimits \left \{\mathbf{b}_N \mathbf{b}^H_N \right\}\right)}.
\end{align}
Then, calculate $\mathop {\mathrm {\mathbb {E}}}\nolimits \left \{\mathbf{b}_k \mathbf{b}^H_k \right\} $
where $\mathbf{b}^H_k = \mathbf{x}^H_k \mathbf{Y}$ as
\begin{align}
\mathop {\mathrm {\mathbb {E}}}\nolimits \left \{\mathbf{b}_k \mathbf{b}^H_k \right\} 
&= \mathop {\mathrm {\mathbb {E}}}\nolimits \left \{\mathbf{Y}^H \mathbf{x}_k \mathbf{x}^H_k \mathbf{Y}\right\}\notag\\
&= \colvec{
  \mathop {\mathrm {\mathbb {E}}}\nolimits \left \{\vert \mathbf{y}^H_1 \mathbf{x}_k \vert^2 \right\} &   &  & \\
   & \mathop {\mathrm {\mathbb {E}}}\nolimits \left \{\vert \mathbf{y}^H_2 \mathbf{x}_k \vert^2 \right\} &  & \\
   &  & \hdots &  \\
   &  &  & \mathop {\mathrm {\mathbb {E}}}\nolimits \left \{\vert \mathbf{y}^H_N \mathbf{x}_k \vert^2 \right\}}.
\end{align}

\subsection{Proof of Theorem \ref{theo_closedform}}\label{AppendixE}
\subsubsection{Compute $\bar{\textbf{D}}_{kk}$}\label{AppendixE1}
\begin{align}\label{hbareq} 
\bar{\textbf{D}}_{kk} &=\mathop {\mathrm {\mathbb {E}}}\nolimits \left \{ \sum_{m=1}^{M}\eta^{1/2}_{mk}\textbf{G}^H_{mk}(\textbf{Y}_{mk}\textbf{A}_{mk})\right \}\notag\\
&= \sqrt{\tau_{\text{u}}\rho_{\text{u}}}\sum_{m=1}^{M}\eta^{1/2}_{mk} \mathop {\mathrm {\mathbb {E}}}\nolimits \left \{ \textbf{G}^H_{mk} \textbf{G}_{mk}\right \}\textbf{A}_{mk}\notag\\
&=L \sqrt{\tau_{\text{u}}\rho_{\text{u}}}\sum_{m=1}^{M}\eta^{1/2}_{mk}\beta_{mk}\textbf{A}_{mk}.
\end{align}
\subsubsection{Compute $\mathbf{\Psi}^b_{kk}$}\label{AppendixE2}
\begin{align}\label{slast} 
\mathbf{\Psi}^b_{kk} =\mathcal{S}_1 - \rho_d\bar{\textbf{D}}_{kk}\bar{\textbf{D}}^H_{kk} + \textbf{I}_{N},
\end{align}
where 
\begin{align}\label{s1last}  
\mathcal{S}_1 &\triangleq \mathbb{E}\left\{\rho\sum_{m=1}^{M}\sum_{n=1}^{M}\sum_{k'=1}^{K}\eta^{1/2}_{mk'}\eta^{1/2}_{nk'}\textbf{G}^H_{mk}\hat{\textbf{G}}_{mk'}\hat{\textbf{G}}^H_{nk'}\textbf{G}_{nk}\right\}\notag\\
&=\mathcal{T}_{1} + \mathcal{T}_{2},
\end{align}
with
\begin{align} 
\mathcal{T}_{1}
&\triangleq \mathop {\mathrm {\mathbb {E}}}\nolimits \left \{\rho\sum_{m=1}^{M}\sum_{k'=1}^{K}\eta_{mk'}\textbf{G}^H_{mk}\textbf{Y}_{mk'}\textbf{A}_{mk'}\textbf{A}^H_{mk'}\textbf{Y}^H_{mk'}\textbf{G}_{mk} \right \},
\end{align}
and
\begin{align} 
\mathcal{T}_{2} &\!\triangleq \mathop {\mathrm {\mathbb {E}}}\nolimits \bigg \{\rho\sum_{m=1}^{M}\sum_{n \ne m}^{M}\sum_{k'=1}^{K}\eta^{1/2}_{mk'}\eta^{1/2}_{nk'}  \textbf{G}^H_{mk} (\sqrt{\tau_{\text{u}}\rho_{\text{u}}}\textbf{G}_{mk}\Phi_{kk'}\textbf{A}_{mk'})\notag\\
&\quad\times(\textbf{A}^H_{nk'}\sqrt{\tau_{\text{u}}\rho_{\text{u}}}\Phi^H_{kk'}\textbf{G}^H_{nk} )\textbf{G}_{nk} \bigg \}.
\end{align}
Firstly, to calculate $\mathcal{T}_{1}$, we have
\begin{align}\label{t1eq} 
\mathcal{T}_{1}&=\rho\sum_{m=1}^{M}\sum_{k'=1}^{K}\eta_{mk'}\mathop {\mathrm {\mathbb {E}}}\nolimits \bigg \{\textbf{G}^H_{mk}\!\left(\sum_{i=1}^{K}\sqrt{\tau_{\text{u}}\rho_{\text{u}}}\textbf{G}_{mi}\Phi_{ik'} + \textbf{W}_{nk'}\!\right)\notag\\
&\quad\times\textbf{A}_{mk'}\textbf{A}^H_{mk'}\left(\sum_{i=1}^{K}\sqrt{\tau_{\text{u}}\rho_{\text{u}}}\Phi^H_{ik'}\textbf{G}^H_{mi} + \textbf{W}^H_{nk'}\right)\textbf{G}_{mk} \bigg \}\notag\\
&=\rho\sum_{m=1}^{M}\sum_{k'=1}^{K}\eta_{mk'}\left\{\tau_{\text{u}}\rho_{\text{u}}(\mathcal{T}_{11}+\mathcal{T}_{12})  + \mathcal{T}_{13}\right\},
\end{align}
where 
\begin{align} 
\mathcal{T}_{11}\label{t11eq} &\triangleq \sum_{i \ne k}^{K}\mathop {\mathrm {\mathbb {E}}}\nolimits \left \{\textbf{G}^H_{mk}\textbf{G}_{mi}\Phi_{ik'}\textbf{A}_{mk'}\textbf{A}^H_{mk'}\Phi^H_{ik'}\textbf{G}^H_{mi}\textbf{G}_{mk}\right\}\notag\\
&\mathop = \limits ^{(b1)}L\sum_{i \ne k}^{K}\beta_{mi}\beta_{mk} \Tr(\Phi_{ik'}\textbf{A}_{mk'}\textbf{A}^H_{mk'}\Phi^H_{ik'})\textbf{I}_{N},\\
\mathcal{T}_{12}\label{t12eq}
&\triangleq \mathop {\mathrm {\mathbb {E}}}\nolimits \left \{\textbf{G}^H_{mk}\textbf{G}_{mk}\Phi_{kk'}\textbf{A}_{mk'}\textbf{A}^H_{mk'}\Phi^H_{kk'}\textbf{G}^H_{mk}\textbf{G}_{mk}\right\}\notag\\
&\mathop = \limits ^{(b2)}L\beta^2_{mk} \textbf{C}_{mkk'},
\end{align}
where $\textbf{C}_{mkk'}$ is a $N \times N$ diagonal matrix that the element $c_{ii} = \sum_{n = 1}^{N} b_{nn} + Lb_{ii}$ with $b$ is an element of $\textbf{B}_{mkk'}= \Phi_{kk'}\textbf{A}_{mk'}\textbf{A}^H_{mk'}\Phi^H_{kk'}$.
and
\begin{align} 
\mathcal{T}_{13}\label{t13eq}
&\triangleq \mathop {\mathrm {\mathbb {E}}}\nolimits \left \{\textbf{G}^H_{mk}\textbf{W}_{nk'}\textbf{A}_{mk'}\textbf{A}^H_{mk'}\textbf{W}^H_{nk'}\textbf{G}_{mk}\right\}\notag\\
&\mathop = \limits ^{(b3)}L\beta_{mk}\Tr(\textbf{A}_{mk'}\textbf{A}^H_{mk'})\textbf{I}_{N},
\end{align}
where $(b1)$, $(b2)$ and $(b3)$ are derived by Lemma \ref{lemma3}.\\  
Substituting (\ref{t11eq}) (\ref{t12eq}) (\ref{t13eq}) into (\ref{t1eq}), we have
\begin{align}\label{t1last} 
&\mathcal{T}_{1}=L\rho\sum_{m=1}^{M}\sum_{k'=1}^{K}\beta_{mk}\eta_{mk'} \bigg\{\tau_{\text{u}}\rho_{\text{u}}\bigg(\sum_{i \ne k}^{K}\beta_{mi}\times\notag\\
&\Tr(\Phi_{ik'}\textbf{A}_{mk'}\textbf{A}^H_{mk'}\Phi^H_{ik'})+\beta_{mk} \textbf{C}_{mkk'}\bigg)  +\Tr(\textbf{A}_{mk'}\textbf{A}^H_{mk'})\bigg\}\textbf{I}_{N}.
\end{align}
To calculate $\mathcal{T}_{2}$, we have
\begin{align}\label{t2last}  
&\mathcal{T}_{2}\notag\\ &=\tau_{\text{u}}\rho_{\text{u}}\rho\sum_{m=1}^{M}\sum_{n \ne m}^{M}\sum_{k'=1}^{K}\eta^{1/2}_{mk'}\eta^{1/2}_{nk'}\notag\\
&\quad\times\mathop {\mathrm {\mathbb {E}}}\nolimits \left \{\textbf{G}^H_{mk}\textbf{G}_{mk}\Phi_{kk'}\textbf{A}_{mk'}\textbf{A}^H_{nk'}\Phi^H_{kk'}\textbf{G}^H_{nk} \textbf{G}_{nk} \right \}\notag\\
&=\tau_{\text{u}}\rho_{\text{u}}\rho\sum_{m=1}^{M}\sum_{n \ne m}^{M}\sum_{k'=1}^{K}\eta^{1/2}_{mk'}\eta^{1/2}_{nk'}\notag\\
&\quad\times\mathop {\mathrm {\mathbb {E}}}\nolimits \left \{\textbf{G}^H_{mk}\textbf{G}_{mk}\right \} \Phi_{kk'}\textbf{A}_{mk'}\textbf{A}^H_{nk'}\Phi^H_{kk'} \mathop {\mathrm {\mathbb {E}}}\nolimits \left \{\textbf{G}^H_{nk} \textbf{G}_{nk} \right \}\notag\\ 
&=\!L^2\tau_{\text{u}}\rho_{\text{u}}\rho\!\!\sum_{m=1}^{M}\!\sum_{n \ne m}^{M}\!\sum_{k'=1}^{K}\eta^{1/2}_{mk'}\!\eta^{1/2}_{nk'} \beta_{mk}\beta_{nk} \Phi_{kk'}\textbf{A}_{mk'}\textbf{A}^H_{nk'}\Phi^H_{kk'}.
\end{align}
The substitution of (\ref{t1last}) (\ref{t2last}) and (\ref{hbareq}) into (\ref{slast}) yields (\ref{seq}). 

\subsection{Proof of Lemma \ref{lemma3}}\label{Appendix_proflemma3} 
Applying the Lindeberg-L$\acute{\text{e}}$vy, we obtain
\begin{align} 
d_{kk,ij} &=\sum_{m=1}^{M}\eta_{mk}^{1/2}\textbf{g}^H_{mk, i}\hat{\textbf{g}}_{mk, j} \overset{d}{\rightarrow} \mathcal{CN}(0, \xi_{kk,j}),\notag\\
&\quad\quad\text{as}\ M\rightarrow\infty, \text{ and } \forall i \neq j,\\
d_{kk,ii}&=\sum_{m=1}^{M}\eta_{mk}^{1/2}\textbf{g}^H_{mk, i}\hat{\textbf{g}}_{mk, i},\notag\\ 
&= \sum_{m=1}^{M}\eta_{mk}^{1/2}\hat{\textbf{g}}^H_{mk, i}\hat{\textbf{g}}_{mk, i} + \sum_{m=1}^{M}\eta_{mk}^{1/2}\tilde{\textbf{g}}^H_{mk, i}\hat{\textbf{g}}_{mk, i},\notag\\ 
&\approx\sum_{m=1}^{M}\eta_{mk}^{1/2}\hat{\textbf{g}}^H_{mk, i}\hat{\textbf{g}}_{mk, i},\notag\\ 
&\overset{d}{\rightarrow} \mathcal{N}\left(L\sum_{m=1}^{M}\sqrt{\eta_{mk}}\gamma_{mk,i},L\sum_{m=1}^{M}\eta_{mk}\gamma_{mk,i}^{2}\right),\notag\\
&\quad\quad\text{as}\ M\rightarrow\infty, \\
d_{kk',ij} &=\sum_{m=1}^{M}\eta_{mk'}^{1/2}\textbf{g}^H_{mk, i}\hat{\textbf{g}}_{mk', j}\overset{d}{\rightarrow} \mathcal{CN}(0, \xi_{kk',j}),\notag\\
&\quad\text{as}\ M\rightarrow\infty,\forall k' \ne k,  \text{ and } i, j = 1,2, \ldots, N. 
\end{align}
The Gaussian approximation is verified by numerical results in the section \ref{Numsec}.

\subsection{Proof of Theorem \ref{theo_SE_protocol2_UC}}\label{Appendix_proftheo_SE_protocol2_UC} 
Denote $\hat{\textbf{D}}_{kk'}$ are MMSE estimated of $\textbf{D}_{kk'}$, and $\tilde{\textbf{D}}_{kk'} \triangleq \textbf{D}_{kk'} - \hat{\textbf{D}}_{kk'}$. 
With assumption that the elements of the effective channels $\textbf{D}_{kk'}$ are Gaussian distributed, the corresponding elements of $\hat{\textbf{D}}_{kk'}$ and $\tilde{\textbf{D}}_{kk'}$ are independent in distribution. Then, with $\mathbf{\Theta}_{k} = \{\hat{\textbf{D}}_{k1}, \hat{\textbf{D}}_{k2}, \ldots , \hat{\textbf{D}}_{kK}\}$, the achievable downlink SE of the $k$-th user in (\ref{R_k1}) when using MMSE-SIC detectors at the receivers can be rewritten as 
\begin{align}\label{}  
R^{\text{Et-SIC}}_k \!=\! \left(\!1 \!-\! \frac{\tau_{\text{u}} + \tau_{\text{d}}}{\tau_c}\!\right)\! \mathbb {E}\! \left \{ \log_{2} \left|\textbf{I}_{N} + \rho\hat{\textbf{D}}^H_{kk} \left(\mathbf{\Psi}^{c'}_{kk}\right)^{-1} \hat{\textbf{D}}_{kk} \right| \right \},
\end{align}
where
\begin{align}\label{psi_c}
\mathbf{\Psi}^{c'}_{kk} = \rho\sum^{K}_{k'\neq k}\hat{\textbf{D}}_{kk'}\hat{\textbf{D}}^H_{kk'} + \rho \sum^{K}_{k'=1}\mathbb {E} \left \{\tilde{\textbf{D}}_{kk'} \tilde{\textbf{D}}^{H}_{kk'} \right\} +  \textbf{I}_{N}.
\end{align}
Next, we compute $\mathbb {E} \left \{\tilde{\textbf{D}}_{kk'} \tilde{\textbf{D}}^{H}_{kk'} \right\}$:
\begin{align}\label{exp_Dkk} 
&\mathbb {E} \left \{\tilde{\textbf{D}}_{kk'} \tilde{\textbf{D}}^{H}_{kk'} \right\}\notag\\
& =\colvec{\mathbb {E} \left \{\tilde{\textbf{d}}^{T}_{kk',1} \tilde{\textbf{d}}^{*}_{kk',1}\right\} & \mathbb {E} \left \{\tilde{\textbf{d}}^{T}_{kk',1} \tilde{\textbf{d}}^{*}_{kk',2}\right\}  & \hdots & \mathbb {E} \left \{\tilde{\textbf{d}}^{T}_{kk',1} \tilde{\textbf{d}}^{*}_{kk',N}\right\}  \\
\mathbb {E} \left \{\tilde{\textbf{d}}^{T}_{kk',2} \tilde{\textbf{d}}^{*}_{kk',1}\right\} & \mathbb {E} \left \{\tilde{\textbf{d}}^{T}_{kk',2} \tilde{\textbf{d}}^{*}_{kk',2}\right\} & \hdots & \mathbb {E} \left \{\tilde{\textbf{d}}^{T}_{kk',2} \tilde{\textbf{d}}^{*}_{kk',N}\right\}\\
\vdots
		& & \ddots  &\\
   \mathbb {E} \left \{\tilde{\textbf{d}}^{T}_{kk',N} \tilde{\textbf{d}}^{*}_{kk',1}\right\} &  &  & \mathbb {E} \left \{\tilde{\textbf{d}}^{T}_{kk',N} \tilde{\textbf{d}}^{*}_{kk',N}\right\}} \notag\\
& =\colvec{\sum^{N}_{j=1} \text{var}(\tilde{d}_{kk',1j}) & 0  & \hdots & 0  \\
0 & \sum^{N}_{j=1} \text{var}(\tilde{d}_{kk',2j})  & \hdots & 0\\
\vdots
		& & \ddots  &\\
   0 & 0 & \hdots & \sum^{N}_{j=1} \text{var}(\tilde{d}_{kk',Nj}) }. 
\end{align}
 Then,  the variance of the MMSE  estimation error of $d_{kk',ij}$ can be calculated as \cite{kay1993fundamentals}
\begin{align} 
\text{var}(\tilde{d}_{kk',ij}) &= \textbf{C}_{d_{kk',ij},d_{kk,ij}} -\textbf{C}_{d_{kk',ij},y_{\text{dp},k,ij}}\notag\\
&\quad\times\textbf{C}^{-1}_{y_{\text{dp},k,ij},y_{\text{dp},k,ij}}\textbf{C}_{y_{\text{dp},k,ij},d_{kk',ij}}.
\end{align}
Following the similar method on section \ref{AppendixB1}, we have
\begin{equation}\label{var_dkkij}
\text{var}(\tilde{d}_{kk',ij}) =
\begin{cases}
\frac{\xi_{kk,i} + \kappa_{kk,i}^2}{\tau_{\text{d}}\rho_{\text{d,p}}(\xi_{kk,i} + \kappa_{kk,i}^2) + 1} & \text{if } k = k',\text{and } i = j \\
\frac{\xi_{kk',j}}{\tau_{\text{d}}\rho_{\text{d,p}}\xi_{kk',j} + 1}  & \text{otherwise }.
\end{cases}
\end{equation}
Finally, substituting (\ref{var_dkkij}) into (\ref{exp_Dkk}), and then plugging (\ref{exp_Dkk}) into (\ref{psi_c}), we obtain (\ref{psi_ckk_final}).

\end{document}